\documentclass[preprint,superscriptaddress,amsmath,amssymb,aps,prx]{revtex4-1}

\usepackage[dvipdfmx]{graphicx,color}
\usepackage{dcolumn}
\usepackage{bm}

\usepackage{comment}
\usepackage{color}

\newcommand{\Part}[3]{ \frac{ \partial^{#3} #1 }{ \partial #2^{#3} } }
\newcommand{\V}[1]{\bm{#1} } 
\newcommand{\Tr}[1]{ \mathop{\rm Tr}_{ #1 } }

\newcommand{\lb}{\left(}
\newcommand{\rb}{\right)}
\newcommand{\lbb}{\left\{}
\newcommand{\rbb}{\right\}}
\newcommand{\lsb}{ \left[ }
\newcommand{\rsb}{ \right] }

\newcommand{\BReq}[1]{(\ref{eq:#1})}
\newcommand{\NReq}[1]{(\ref{eq:#1})}

\newcommand{\BReqs}[2]{(\ref{eq:#1}, \ref{eq:#2})}

\newcommand{\BReqss}[2]{(\ref{eq:#1}--\ref{eq:#2})}
\newcommand{\Rfig}[1]{figure\ \ref{fig:#1}}

\newcommand{\Lfig}[1]{\label{fig:#1}}
\newcommand{\Leq}[1]{\label{eq:#1}}

\newcommand{\Lsec}[1]{\label{sec:#1}}
\newcommand{\Rapp}[1]{Appendix\ \ref{sec:#1}}
\newcommand{\be}{\begin{eqnarray}}
\newcommand{\ee}{\end{eqnarray}}
\newcommand{\ba}{\begin{array}}
\newcommand{\ea}{\end{array}}
\newcommand{\no}{\nonumber}

\newcommand{\subbe}{\begin{subequations}}
\newcommand{\subee}{\end{subequations}}
\DeclareMathOperator*{\argmin}{arg\,min}

\begin{document}

\preprint{APS/123-QED}

\title{Sparse approximation based on a random overcomplete basis}

\author{Yoshinori Nakanishi-Ohno}
\email{nakanishi@mns.k.u-tokyo.ac.jp}
\affiliation{Graduate School of Frontier Sciences, The University of Tokyo, Kashiwa, Chiba, 277-8561, Japan}
\affiliation{Research Fellow of Japan Society for the Promotion of Science, Chiyoda, Tokyo, 102-0083, Japan}
\author{Tomoyuki Obuchi}
\email{obuchi@sp.dis.titech.ac.jp}
\affiliation{Interdisciplinary Graduate School of Science and Engineering, Tokyo Institute of Technology, Yokohama, Kanagawa, 226-8502, Japan}
\author{Masato Okada}
\email{okada@k.u-tokyo.ac.jp}
\affiliation{Graduate School of Frontier Sciences, The University of Tokyo, Kashiwa, Chiba, 277-8561, Japan}
\author{Yoshiyuki Kabashima}
\email{kaba@dis.titech.ac.jp}
\affiliation{Interdisciplinary Graduate School of Science and Engineering, Tokyo Institute of Technology, Yokohama, Kanagawa, 226-8502, Japan}

\date{\today}

\begin{abstract}
We discuss a strategy of sparse approximation that is based on the use of an overcomplete basis, and evaluate its performance when a random matrix is used as this basis. 
A small combination of basis vectors is chosen from a given overcomplete basis, according to a given compression rate, such that they compactly represent the target data with as small a distortion as possible. 
As a selection method, we study the $\ell_0$- and $\ell_1$-based methods, which employ the exhaustive search and $\ell_1$-norm regularization techniques, respectively. 
The performance is assessed in terms of the trade-off relation between the representation distortion and the compression rate. 
First, we evaluate the performance analytically in the case that the methods are carried out ideally, using methods of statistical mechanics. 
The analytical result is then confirmed by performing numerical experiments on finite size systems, and extrapolating the results to the infinite-size limit. 
Our result clarifies the fact that the $\ell_0$-based method greatly outperforms the $\ell_1$-based one. 
An interesting outcome of our analysis is that any small value of distortion is achievable for any fixed compression rate $r$ in the large-size limit of the overcomplete basis, for both the $\ell_0$- and $\ell_1$-based methods. 
The difference between these two methods is manifested in the size of the overcomplete basis that is required in order to achieve the desired value for the distortion. As the desired distortion decreases, the required size grows in a polynomial and an exponential manners for the $\ell_0$- and $\ell_1$-based methods, respectively. 
Second, we examine the practical performances of two well-known algorithms, orthogonal matching pursuit and approximate message passing, when they are used to execute the $\ell_0$- and $\ell_1$-based methods, respectively. 
Our examination shows that orthogonal matching pursuit achieves a much better performance than the exact execution of the $\ell_1$-based method, as well as approximate message passing. 
However, regarding the $\ell_0$-based method, there is still room to design more effective greedy algorithms than orthogonal matching pursuit. Finally, we evaluate the performances of the algorithms when they are applied to image data compression.
\end{abstract}

\pacs{Valid PACS appear here}
\maketitle

\section{Introduction}
Information processing based on the sparseness of various data is an active area of research.
This sparseness means that data are typically expressed by a small combination of non-zero components when a proper basis is used.
The significance of sparseness for information processing had already begun to be noted when principal component analysis was invented, in 1901 \cite{Pearson1901}.
Low-rank approximation of a matrix is known to be a useful method of collaborative filtering for recommendation systems \cite{Eckart36,Su09,Markovsky12}.
In neuroscience, the sparse-coding hypothesis has gradually been accepted as a method of elucidating visual and auditory systems \cite{Olshausen96,Olshausen97,Olshausen04,Terashima09,Terashima12,Terashima13}.
Recent interest in information processing with sparse data has been triggered by compressed sensing, since it was demonstrated that $\ell_1$-norm minimization can give exact solutions in a reasonable time, under appropriate conditions \cite{Donoho06a,Candes05,Candes06a,Candes06b}.

In this study, we discuss sparse data processing from a different viewpoint, namely that of sparse approximation.
Sparse approximation refers to the process of representing target data by a small number of non-zero elements, 
the purpose of which is to achieve a better trade-off relation between the representation distortion and the compression rate \cite{Natarajan95,Davis97,Temlyakov98,Temlyakov99,Temlyakov00,Temlyakov03,Gilbert03,Tropp03,Tropp04,Donoho06b}.
We adopt a strategy of sparse approximation that utilizes an overcomplete basis (OCB).
An OCB can also be called a frame in the field of signal processing.
OCBs contain more basis vectors than the dimension of target data.
This means that a better and smaller set of basis vectors may be chosen to compactly express the data.
Therefore, in terms of the trade-off relation, the OCB-based strategy is expected to outperform naive strategies such as random projection.

For selecting basis vectors from an overcomplete basis, we discuss the $\ell_0$- and $\ell_1$-based methods, which employ the exhaustive search and $\ell_1$-norm regularization techniques, respectively.
Our adoption of these methods is motivated by their application in compressed sensing \cite{Foucart13,Rish14}.
Focusing on the trade-off relation, we evaluate the performance of sparse approximation from two different viewpoints. First, we theoretically analyze the ideal performance that is achieved when the $l_0$- and $l_1$-based methods are performed exactly, by using methods of statistical mechanics.
We regard the distortion and the compression rate as the thermal averages of physical quantities derived from partition functions.
In the large-system limit, these are assessed by the replica method and the saddle-point method \cite{Nishimori01,Dotsenko01}.
In order to validate the results of our analysis, we extrapolate physical quantities in the limit, from finite-size results obtained using the exchange Monte Carlo method \cite{Swendsen86,Hukushima96} and quadratic programming.
Second, we investigate the practical performance of the OCB-based strategy.
We examine the performances of two well-known algorithms, orthogonal matching pursuit \cite{Pati93,Davis94} and approximate message passing \cite{Donoho09}, when they are employed to approximately execute the $\ell_0$- and $\ell_1$-based methods, respectively.
We also apply the approximate algorithms to a task of image data compression and evaluate their performances, as a practical example.

The rest of this paper is organized as follows.
In section \ref{setting}, we set up the problem of sparse approximation that we will focus on, and explain the $\ell_0$- and $\ell_1$-based methods and related work.
In section \ref{effect}, we analyze the ideal performances of these methods, in terms of the trade-off relation.
In section \ref{practice}, we discuss the practical performance of the OCB-based strategy, and its application to image data.
In section \ref{conclusion}, we conclude this paper.

\section{Problem setting}\label{setting}

\subsection{Sparse approximation using a random overcomplete basis}
Given a data vector $\bm{y}\in\mathbb{R}^{M}$ and a compression rate $r$, the purpose of sparse approximation is to obtain a compressed representation $\bm{x}\in\mathbb{R}^{N}$ using a basis matrix $\bm{A}=(\bm{a}_1,\dots,\bm{a}_N)\in\mathbb{R}^{M\times N}$, while keeping the representation distortion $\epsilon$ as small as possible. The compression rate $r$ is defined as the ratio of the number of non-zero components of $\V{x}$ to the dimension of the data vector. That is,
\begin{eqnarray}
	r=\frac{||\bm{x}||_0}{M},
\end{eqnarray}
where $||\cdot||_0$ denotes the so-called $\ell_0$-norm of a vector. The $\ell_0$-norm represents the number of non-zero elements of a vector, defined as $||\bm{v}||_0=\sum_i|v_i|_0$, where $|v_i|_0$ is equal to $0$ ($v_i=0$) or $1$ ($v_i\not=0$). We measure the distortion using the mean squared error, as
\begin{eqnarray}
	\epsilon&=&\frac{1}{2M}||\bm{y}-\bm{Ax}||_2^2,
\Leq{epsilon}
\end{eqnarray}
where $||\cdot||_2$ is the $\ell_2$-norm of a vector, defined as $||\bm{v}||_2=\sqrt{\sum_iv_i^2}$. 
Note that this representation distortion measures how close a data vector $\bm{y}$ is described by a sparse representation $\bm{x}$ with a given basis $\bm{A}$, 
and it is different from the reconstruction error often used to measure the distance between an original sparse signal $\bm{x}_0$ and an estimated sparse representation $\hat{\bm{x}}$ in the field of signal processing and compressed sensing.
For our purpose of an analytical evaluation of $\epsilon$, we consider the case where the elements of the data vector $\bm{y}$ are independently and identically distributed (i.i.d.) random variables from the normal distribution, whose mean and variance are 0 and $\sigma_y^2$, respectively, and together are denoted by $\mathcal{N}(0,\sigma_y^2)$. The elements of the basis matrix $\bm{A}$ are also i.i.d. random variables from $\mathcal{N}(0,M^{-1})$. Then, the matrix $\V{A}$ is almost surely of rank $\min(M,N)$, and the distortion becomes a random variable. 

\begin{figure}[htbp]
\begin{tabular}{cc}
\begin{minipage}{0.5\hsize}
\begin{center}
\includegraphics[width=\hsize]{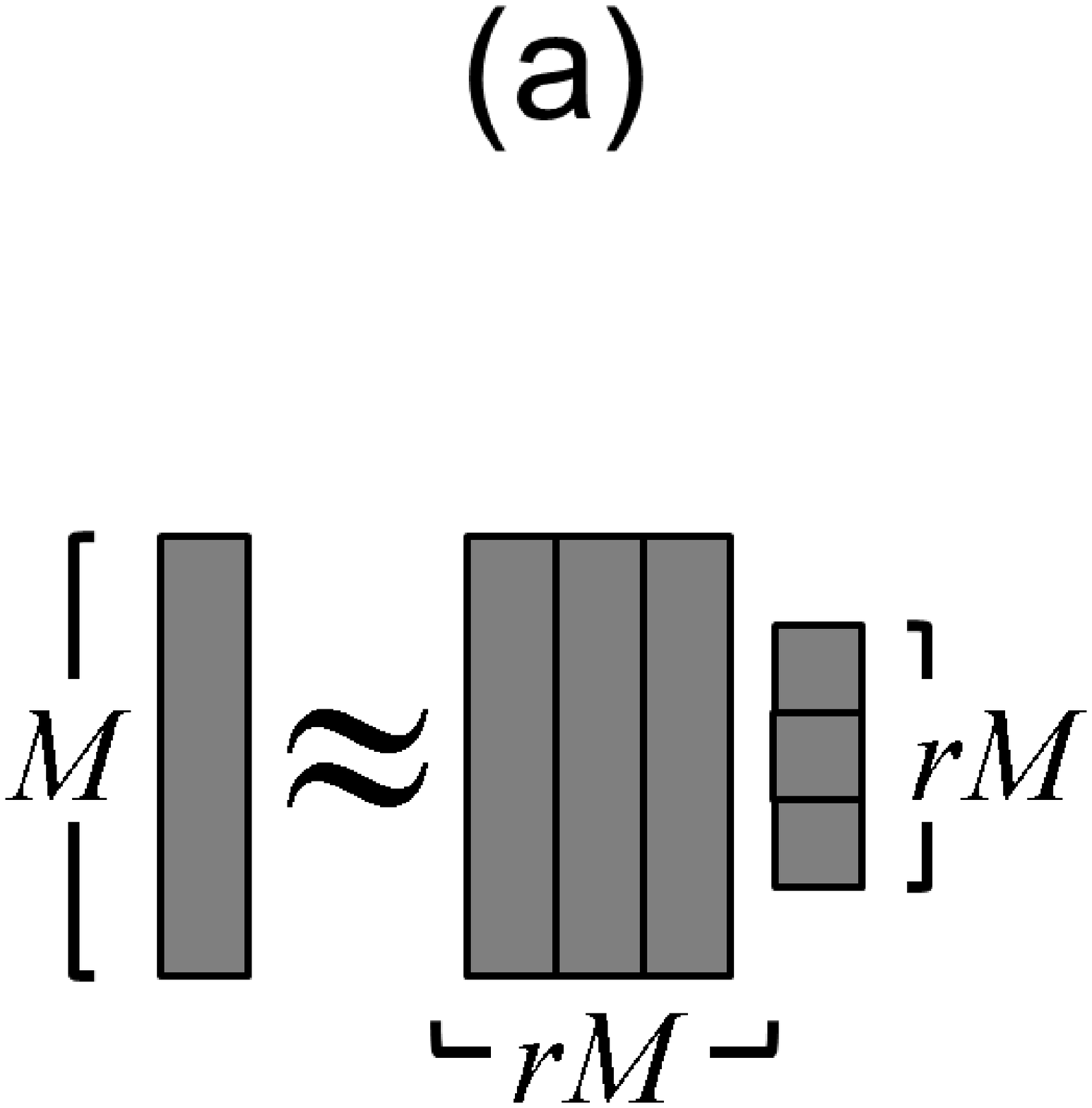}
\end{center}
\end{minipage}
\begin{minipage}{0.5\hsize}
\begin{center}
\includegraphics[width=\hsize]{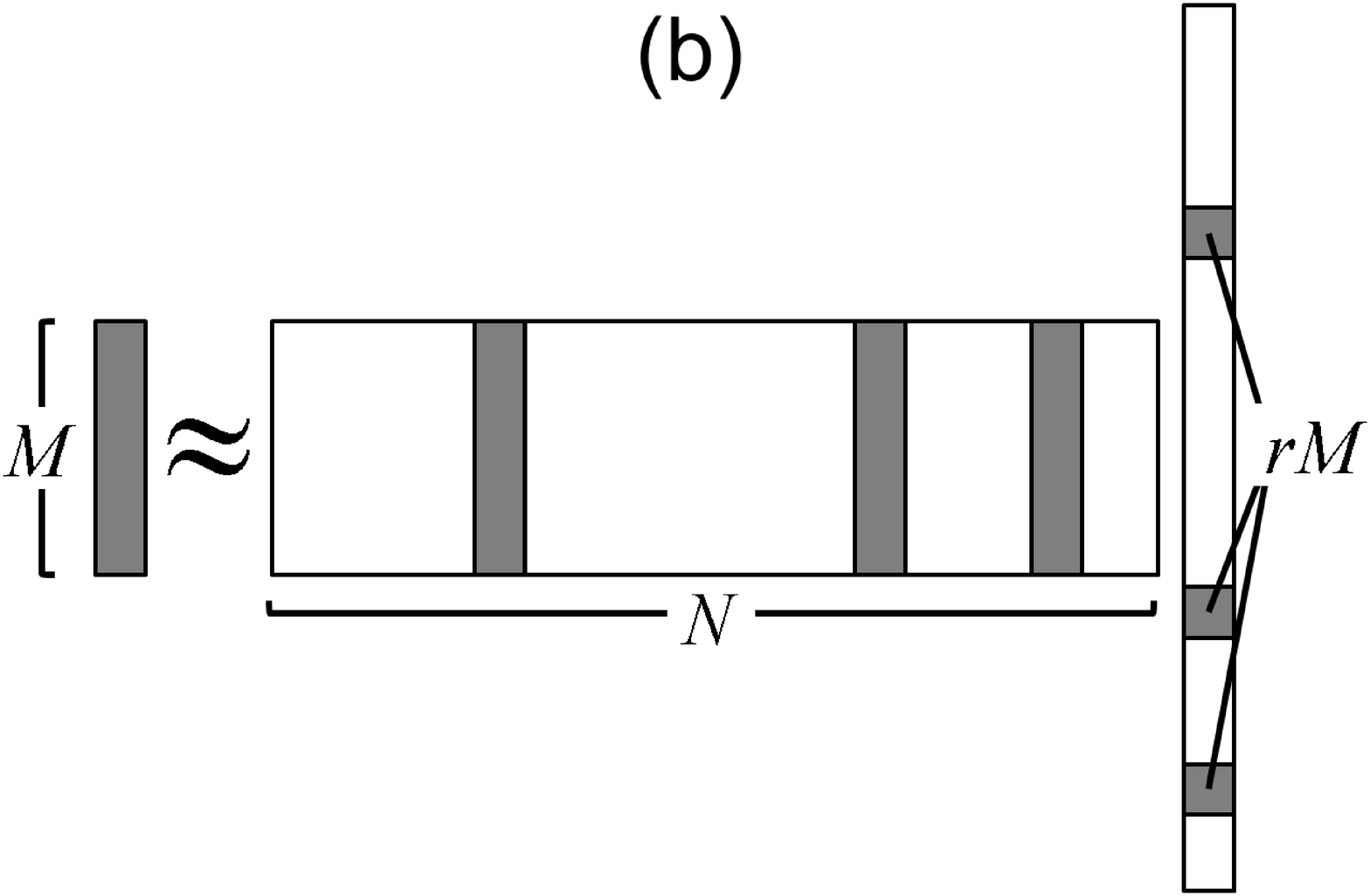}
\end{center}
\end{minipage}
\end{tabular}
\caption{Schematic diagrams of sparse approximation. (a) Naive method. (b) OCB-based strategy.}
\label{DRconcept}
\end{figure}

If $N=rM$, the minimization of \BReq{epsilon} is nothing but the method of least squares (LS), and the corresponding compressed vector is easily obtained as
\be
\hat{\V{x}}=\V{A}^{+}\V{y},
\ee
where $\bm{A}^+$ is the pseudoinverse of $\V{A}$, given by
\be
\bm{A}^+=(\bm{A}^{\rm T}\bm{A})^{-1}\bm{A}^{\rm T}.
\ee
Let us call this the naive method, which is illustrated in figure \ref{DRconcept}~(a). In the large-size limit $M\to \infty$, the corresponding distortion converges to 
\begin{eqnarray}
	\epsilon_{\rm naive}&=&\frac{1-r}{2}\sigma_y^2,
\Leq{epsilon_naive}
\end{eqnarray}
with probability one. In general, in the limit $M\to \infty$ certain random variables, such as $\epsilon$, have the so-called self-averaging property, and will almost surely converge to their average values. This enables us to present a clear discussion, and hereafter we focus on this limit.  

On the other hand, for $N> rM$ we have a lot of options in choosing a combination of $rM$ basis vectors from the matrix, as illustrated in figure \ref{DRconcept}~(b). If the chosen combination is more suitable for representing the data vector than one that is chosen randomly, then the distortion becomes smaller than $\epsilon_{\rm naive}$. This is the idea behind the OCB-based strategy. However, this strategy presents the problem of how to choose the combination of basis vectors. We investigate the performances of $\ell_0$- and $\ell_1$-based methods.

\subsection{Methods}
\subsubsection{$\ell_0$-based method}
The basic idea of the $\ell_0$-based method is to minimize the distortion by choosing the best combination of $rM$ basis (column) vectors from a given OCB. More generally, we would like to define the distortion as a function of the chosen combination of basis vectors, and to control it in a simple manner. This motivates us to introduce a binary vector $\bm{c}\in\{1,0\}^N$, to store information on whether each basis vector is chosen ($c_i=1$) or not ($c_i=0$). We also introduce a distortion, labelled by $\V{c}$, with
\begin{eqnarray}
\epsilon\lb \V{c} | \V{y},\V{A} \rb=
\min_{\bm{x}}
\lbb
\frac{1}{2M}||\bm{y}-\bm{A}(\bm{c}\circ\bm{x})||_2^2
\rbb
,
\end{eqnarray}
where $\circ$ is the Hadamard product of two vectors, defined as $(\bm{v}\circ\bm{w})_i=v_iw_i$. In addition, we define an entropy function $s(\epsilon| \V{y},\V{A})$ to represent the number of configurations $\V{c}$ that give a value of $\epsilon$ for the distortion, as follows:
\begin{eqnarray}
	s(\epsilon | \V{y},\V{A})
	=\frac{1}{M}\ln 
	\lb 	
	\#\{\bm{c}\ |\ ||\bm{c}||_0=rM\wedge \epsilon(\bm{c}| \V{y},\V{A}) =\epsilon \}
	\rb	,
	\label{entropydef}
\end{eqnarray}
where $\#$ denotes the number of elements of the following set. 

This entropy function is expected to be analytic and convex upward with respect to $\epsilon$, and cannot be negative, by definition. A typical shape of the entropy is depicted in \Rfig{entropyshape}.
\begin{figure}[htbp]
\begin{center}
\includegraphics[height=0.25\hsize,width=0.4\hsize]{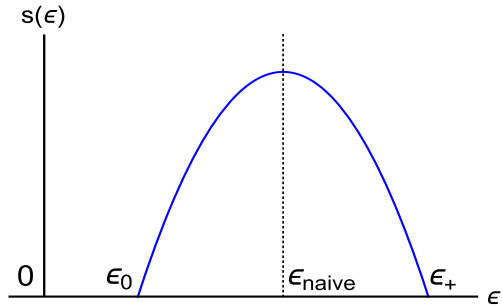}
\caption{A schematic shape of the entropy function. The smaller zero point of the entropy, $\epsilon_0$, corresponds to the minimum of the representation distortion connected to the best combination of the basis vectors, and the point giving the largest entropy value is $\epsilon_{\rm naive}$ because random choice is considered to select one of the most typical combinations of basis vectors. }
\Lfig{entropyshape}
\end{center}
\end{figure}
There are two zero points in the entropy function and the smaller and larger ones are denoted by $\epsilon_0$ and $\epsilon_+$, respectively.
The smaller zero point $\epsilon_0$ of the entropy function, $s(\epsilon_0)=0$, gives the minimum value of the distortion  
\begin{eqnarray}
\epsilon_{0}\lb \V{y},\V{A} \rb
=\min_{\V{c}} \epsilon\lb \V{c} | \V{y},\V{A} \rb
\ \ \ \textrm{subj. to}\ ||\bm{c}||_0= rM.
\end{eqnarray}
Hence, our original motivation for introducing the $\ell_0$-based method, to find the minimum distortion led by the best combination of basis vectors, can be achieved through the evaluation of the entropy function. In addition, the evaluation of the entropy function is easier than the direct evaluation of $\epsilon_0$, and moreover the entropy function provides more information about the space of the variables $\V{c}$, which can be useful for practical applications such as designing algorithms. Thus, the entropy function $s(\epsilon)$ is the primary object of our analysis in the $\ell_0$-based method. A similar analysis has been proposed for examining the weight space structure of multilayer perceptrons~\cite{Monasson94}.

\subsubsection{$\ell_1$-based method}
The $\ell_0$-based method is the most closely matched to the original idea of the OCB-based strategy. However, its algorithmic realization of searching combinations of basis vectors is computationally inefficient, because it requires an exponentially growing computational cost as the system size $N$ increases. In practical situations, instead of the $\ell_0$-based method, a method based on $\ell_1$-norm regularization can be employed. This motivates us to examine the following $\ell_1$-based method. 

Our $\ell_1$-based method arises from the following minimization problem:
\begin{eqnarray}
\hat{\V{\xi}}=\argmin_{\bm{\xi}}
\lbb 
\frac{1}{2}||\bm{y}-\bm{A\xi}||_2^2+\lambda||\bm{\xi}||_1
\rbb
,\label{l1regdef}
\end{eqnarray}
where $||\cdot||_1$ is the $\ell_1$-norm of a vector, defined as $||\bm{v}||_1=\sum_i|v_i|$, with the absolute value denoted by $|\cdot|$. The solution of this minimization problem, $\hat{\V{\xi}}$, provides useful information for finding the compressed vector we desire. This minimization problem is equivalent to the least absolute shrinkage and selection operator, also known as LASSO \cite{Tibshirani96}.
The main benefit of this approach represented by (\ref{l1regdef}) is the computational ease of performing the minimization. As the objective function of (\ref{l1regdef}) is convex, its minimization can be exactly carried out with a computational time in $O(N^3)$, using versatile algorithms of quadratic programming. Furthermore, the $\ell_1$-norm term in (\ref{l1regdef}) results in a sparsifying effect in $\hat{\V{\xi}}$, and its coefficient $\lambda$ is adjusted according to the compression rate. Namely, $\lambda$ is chosen so that $||\hat{\V{\xi}}||_0=rM$. 

Our aim in the analysis in the $\ell_1$-case is to evaluate the distortion resulting from $\hat{\V{\xi}}$. The expression of the distortion is given by
\be
\epsilon_1=\frac{1}{2M}|| \V{y}-\V{A}\V{\hat{\xi} } ||_2^2.
\ee
An inconvenience presented by this distortion is that it is not minimized on the set of basis vectors chosen by $\hat{\V{\xi}}$, owing to the presence of the $\ell_1$-norm term. In order to remove this extra distortion, we determine again the values of the non-zero components by purely minimizing the distortion,
after the support estimation of the compressed vector by the $\ell_1$-norm regularization.
This procedure is described as follows:
\begin{eqnarray}
	\epsilon^{\rm LS}_1=\min_{\bm{x}}\lbb \frac{1}{2M}||\bm{y}-\bm{A}(|\bm{\hat{\xi}}|_0\circ\bm{x})||_2^2 \rbb,
	\label{l1pidef}
\end{eqnarray}
where $|\cdot|_0$ of a vector is defined by $(|\bm{v}|_0)_i=|v_i|_0$. 
This can be carried out by the method of LS for 
the sub-matrix of $\bm{A}$ that is composed of columns corresponding to $|\hat{\xi}_i|_0=1$.
These two quantities, $\epsilon_1$ and $\epsilon^{\rm LS}_1$, are the objects of our analysis in the $\ell_1$ case. 

\subsection{Related Work}
The problem of sparse approximation has been studied widely in the fields of signal processing, statistics and information theory.
Sparse approximation involves searching for an optimal small combination of given basis vectors, and it was proved to be NP-hard \cite{Natarajan95}.
In our setting, we seek a linear combination of a given number of basis vectors to approximate a given signal with as small a representation distortion as possible \cite{Tropp04}.
This setting is also called $N$-term approximation \cite{Temlyakov98,Temlyakov99,Cohen09}.
As stated after equation (2), our purpose is to minimize the distortion in describing a given signal by a sparse representation.
Note again that this distortion is different from the reconstruction error used to measure the distance between an original sparse signal and an estimated signal from scarce data in compressed sensing.
Our motivation is similar to that of rate-distortion theory for lossy data compression in information theory \cite{Cover06}.

We investigate the performances of the $\ell_0$- and $\ell_1$-based methods in solving the sparse approximation problem.
In the $\ell_0$ case, the exhaustive search is considered to be an absolute method for obtaining the most suitable representation.
A major contribution of this paper is the theoretical analysis of the exhaustive search method for the sparse approximation problem, by using methods of statistical mechanics.

In order to reduce a computational cost of the $\ell_0$-based method, some greedy algorithms were proposed.
Orthogonal matching pursuit (OMP) is a well-known greedy algorithm \cite{Pati93,Davis94}.
The approximation bounds of OMP was proved and has been improved theoretically by previous studies \cite{Gilbert03,Tropp03,Tropp04,Das08,Das11}.
On the other hand, the $\ell_1$-based method based on convex relaxation is known to be useful such as basis pursuit \cite{Chen98} and LASSO \cite{Tibshirani96}.
The problem of sparse approximation allows a distortion in a compressed representation, though it should be small, and we evaluate the performance of the method of $\ell_1$-norm regularization equivalent to LASSO.
In this paper, we are also interested in comparing greedy algorithms and convex-relaxation approach.

\section{Analysis of ideal performance}\label{effect}

\subsection{Analytical treatment in the limit $M\to\infty$}
We investigate the limit $M\to \infty$, as stated above. For this purpose, we employ some statistical mechanical tools, which provide useful assistance investigating this limit. According to the terminology of statistical mechanics, we call the limit $M\to \infty$ the thermodynamic limit, and the average over $\V{y}$ and $\V{A}$ the configurational average, which is denoted by $\lsb \cdot \rsb_{\V{y},\V{A}}$. In taking the limit $M \to \infty$, the aspect ratio of the basis matrix, $\alpha=M/N$, is fixed. 

\subsubsection{$\ell_0$-based method}
A versatile technique of statistical mechanics is to introduce a generating function $Z$ of an energy function $\mathcal{H}$, called a partition function. This defines a canonical distribution $p$. In the $\ell_0$ case, we define the energy function, partition function, and canonical distribution respectively as follows:
\begin{eqnarray}
&&
\mathcal{H}_0(\bm{c};\beta|\V{y},\V{A})=-\frac{1}{\beta}\ln\int\mathrm{d}_{\bm{c}}\bm{x}\mathrm{e}^{-\frac{\beta}{2}||\bm{y}-\bm{A}(\bm{c}\circ\bm{x})||_2^2},
\label{l0energy}
\\ &&	
Z_0(\mu,\beta |\V{y},\V{A})
=\sum_{\bm{c}}\delta(Mr-||\bm{c}||_0)\mathrm{e}^{-\mu\mathcal{H}_0(\bm{c};\beta|\V{y},\V{A})}
\equiv \Tr{\V{c}} \mathrm{e}^{-\mu\mathcal{H}_0(\bm{c};\beta|\V{y},\V{A})},
\Leq{Z_0}
\\ &&
	p_0(\bm{c};\mu,\beta|\V{y},\V{A})=\frac{1}{Z_0(\mu,\beta|\V{y},\V{A})}\delta(Mr-||\bm{c}||_0)\mathrm{e}^{-\mu\mathcal{H}_0(\bm{c},\beta|\V{y},\V{A})},
\end{eqnarray}
where $\int\mathrm{d}_{\bm{c}}x_i$ is equal to $\int\mathrm{d}x_i$ ($c_i=1$) or $1$ ($c_i=0$). 
The parameter $\beta$ denotes the inverse temperature corresponding to the method of LS, and the limit of $\beta\to+\infty$ will be taken in accordance with the execution of the method.
The parameter $\mu$ denotes the inverse temperature corresponding to the support estimation, and plays an important role in combining the entropy and the distortion values to depict the entropy curve as shown in figure \ref{fig:entropyshape}.
The energy function is related to the distortion of a given basis-vector choice $\bm{c}$ as follows:
\begin{eqnarray}
	\frac{1}{M}\lim_{\beta\to\infty}\mathcal{H}_0(\bm{c};\beta|\V{y},\V{A})&=&
	\epsilon(\V{c}|\V{y},\V{A}).
	\label{PIenergy}
\end{eqnarray}
The cumulant generating function $\phi_{0}(\mu|\V{y},\V{A})$ is obtained from $Z_0$ by
\begin{eqnarray}
	\phi_0(\mu|\V{y},\V{A})&=&\lim_{\beta\to\infty}
	\frac{1}{M}\ln Z_0(\mu,\beta|\V{y},\V{A}),
	\Leq{Z_0 to phi_0}
\end{eqnarray}
and is connected to the entropy (\ref{entropydef}) by the Legendre transformation in the large $M$ limit, as
\be
\phi_{0}(\mu |\V{y},\V{A})=
\max_{\epsilon_{0} \leq \epsilon \leq \epsilon_{+}} 
\lbb s(\epsilon|\V{y},\V{A})-\mu \epsilon \rbb.
\Leq{Legendre_0}
\ee
The maximization problem of \BReq{Legendre_0} must be solved on the well-defined region of $s$, which requires appropriate bounds the minimum value of distortion $\epsilon_{0}$ and the maximum value of distortion $\epsilon_+$. Overall, we can calculate the object of our analysis, $s(\epsilon)$, through the inverse Legendre transformation, once we have obtained $\phi_0$. Therefore, we turn our attention to the calculation of $\phi_0$.  

The cumulant-generating function has the self-averaging property, as does the entropy, and we assess the configurational average, given by
\be
\phi_{0}(\mu)= \lsb \phi_{0}(\mu|\V{y},\V{A}) \rsb_{\V{y},\V{A}}.
\Leq{phi_0-ave}
\ee
We employ the replica method in order to calculate this average, and a detailed analysis is provided in Appendix A. Though it is possible that the correct solution to the $\ell_0$-based method might break replica symmetry (RS), the result under the RS ansatz is given by
\begin{eqnarray}
	&&
	\phi_0(\mu)
	=\mathop{\rm extr}\limits_{\hat{\Theta}_0}
	\Biggl\{
	\frac{1}{2}\ln\frac{1+\chi}{1+\chi+\mu(Q-q)}-\frac{1}{2}\frac{\mu(q+\sigma_y^2)}{1+\chi+\mu(Q-q)}
	\Biggr.
	\no\\&& \hspace{-0.5cm}
	+\frac{1}{2}
	\left(
	\hat{r}r+\hat{Q}Q-\frac{\hat{\chi}}{\mu}\chi+\hat{q}q
	\right)
	\Biggl.+\frac{1}{\alpha}\int\mathrm{D}z\ln(
	1+Y)
	\Biggr\},
	\Leq{phi_0}
\end{eqnarray}
where $\mathop{\rm extr}\limits_{\Theta}\{\cdot\}$ denotes the operation of extremization with respect to $\Theta$, $\hat{\Theta}_0=\{Q,\chi,q,\hat{r},\hat{Q},\hat{\chi},\hat{q}\}$, and $\int\mathrm{D}z=\int\frac{\mathrm{d}z}{\sqrt{2\pi}}\mathrm{e}^{-\frac{z^2}{2}}$, and we set
\be
Y\equiv \sqrt{\frac{ \hat{\chi}+\hat{Q} }{ \hat{Q}+\hat{q}}}
	\mathrm{e}^{-\frac{1}{2}\hat{r}+\frac{1}{2}\frac{\hat{q}}{\hat{Q}+\hat{q}}z^2}.
	\Leq{Y_l0}
\ee
By applying the extremization condition, we obtain the following equations of state (EOSs): 
\subbe
\Leq{EOSs_l0}
\be
&&
\hat{\chi}=\mu^2\lbb \frac{\Delta}{(1+\chi)(1+\chi+\mu \Delta)}
+\frac{\sigma_y^2+q}{(1+\chi+\mu \Delta)^2}\rbb,
\Leq{chihat_l0}
\\ &&
\hat{Q}=\mu \lbb \frac{1}{1+\chi+\mu \Delta}
-\frac{\mu(\sigma_y^2+q)}{(1+\chi+\mu \Delta)^2}\rbb,
\Leq{Qhat_l0}
\\ &&
\hat{q}=\mu^2 \frac{\sigma_y^2+q}{(1+\chi+\mu \Delta)^2},
\Leq{qhat_l0}
\\ &&
r=\frac{1}{\alpha}\int \mathrm{D}z \frac{Y }{1+Y },
\Leq{r_l0}
\\ &&
\chi=\frac{\mu r}{\hat{\chi}+\hat{Q}},
\Leq{chi_l0}
\\ &&
Q=r \frac{\hat{\chi}-\hat{q}}{(\hat{\chi}+\hat{Q})(\hat{Q}+\hat{q})}
+
\frac{1}{\alpha}
\frac{\hat{q}}{(\hat{Q}+\hat{q})^2}\int \mathrm{D}z~ z^2\frac{ Y }{1+Y},
\Leq{Q_l0}
\\ &&
q=
\frac{1}{\alpha}
\frac{\hat{q}}{(\hat{Q}+\hat{q})^2}
\int \mathrm{D}z~ z^2 \lb \frac{ Y }{1+Y} \rb^2.
\Leq{q_l0}
\ee
\subee
where we write $\Delta=Q-q$. From the EOSs, we obtain some simple and general relations, which we summarize here for later convenience:
\subbe
\be
&&
\hat{\chi}+\hat{Q}=\frac{ \mu}{1+\chi}, 
\\ &&
\hat{Q}+\hat{q}=\frac{\mu}{1+\chi+\mu \Delta}, 
\\ &&
\hat{\chi}-\hat{q}=\frac{ \mu^2 \Delta}{(1+\chi)(1+\chi+\mu \Delta)},
\\ &&
\chi=\frac{r}{1-r}.
\Leq{chi-simples}
\ee 
\Leq{simples_l0}
\subee
The relation involving the entropy, \BReq{Legendre_0}, enables us to employ a convenient parametric form of $\epsilon(\mu)$ and $s(\mu)=s(\epsilon(\mu))$, and \BReqs{EOSs_l0}{simples_l0} allow us to simplify $\epsilon(\mu)$, as
\be
&&
\epsilon(\mu)=-\Part{\phi_0(\mu)}{\mu}{}=\frac{\hat{\chi}}{2\mu^2},
\Leq{epsilon(mu)}
\\&&
s(\mu)=\phi_0(\mu)+\mu\epsilon(\mu).
\Leq{s(mu)}
\ee
The explicit form of $s(\mu)$ is not enlightening, and therefore we omit it. 
As the value of $\mu$ is increased from $\mu=0$, the point of $(\epsilon,s)$ moves along the entropy curve from the summit ($\mu=0$) in the direction of decreasing the distortion ($\mu>0$) as shown in figure \ref{fig:entropyshape}.
When the entropy curve crosses the zero-entropy line at $\mu=\mu_0$, the minimum distortion is given by
\begin{eqnarray}
	\epsilon_0=\epsilon(\mu_0).
\end{eqnarray}

Here, we make a technical remark on the derivation of \BReq{phi_0}. In contrast to the usual prescription of the replica method, we require two different replica numbers for the present analysis, because we have two different integration variables, $\V{x}$ and $\V{c}$, in the calculation of $\phi_0$. Using \BReqs{Z_0 to phi_0}{phi_0-ave}, and introducing a variable $\nu=\mu/\beta$, we can rewrite $\phi_0(\mu)$ as 
\be
&&
\phi_{0}(\mu)=
\lim_{\nu \to 0} \frac{1}{M} \lsb \ln \Tr{\V{c}} 
\lb 
\int \mathrm{d}_{\V{c}}\V{x} \mathrm{e}^{-\frac{1}{2}\frac{\mu}{\nu}||\V{y}-\V{A}(\V{c}\circ \V{x}) ||_2^2   }  
\rb^{\nu}
\rsb_{\V{y},\V{A}} 
\no \\ &&
=
\lim_{n \to 0} \lim_{\nu \to 0} 
 \frac{1}{Mn} \ln 
  \lsb
   \lbb
    \Tr{\V{c}} 
      \lb 
        \int \mathrm{d}_{\V{c}}\V{x} \mathrm{e}^{-\frac{1}{2}\frac{\mu}{\nu}||\V{y}-\V{A}(\V{c}\circ \V{x}) ||_2^2   }  
      \rb^{\nu}
   \rbb^n
  \rsb_{\V{y},\V{A}}.
  \Leq{replica_l0}
\ee
In the last line, we use the replica identity $ \lsb \ln X \rsb_{\V{y},\V{A}} =\lim_{n\to0} (1/n)\ln  \lsb X^n \rsb_{\V{y},\V{A}}$. We identify $n$ and $\nu$ as the two replica numbers, and assume that they are natural numbers, which enables us to expand the powers and to calculate the configurational average. The remaining calculations follow the usual procedure of the replica method, and we assume the RS ansatz in the order parameters. Our present framework in calculating $\phi_0$ is actually similar to the one-step replica-symmetry-breaking (1RSB) ansatz. In this identification, $\nu$ is identified as the 1RSB breaking parameter (usually written as $m$), and each configuration of $\V{c}$ corresponds to a pure state in the 1RSB free-energy landscape; the entropy can be regarded as complexity. The analytical results obtained on the basis of RS assumption will be justified later, in a comparison with numerical calculations.

\subsubsection{$\ell_1$-based method}

\paragraph{Derivation of $\epsilon_1$}
Similarly to the case of the $\ell_0$-based method, the energy function, partition function, and canonical distribution of the $\ell_1$ case are defined respectively as
\begin{eqnarray}
&&
	\mathcal{H}_1(\bm{\xi}|\V{y},\V{A})=\frac{1}{2}||\bm{y}-\bm{A\xi}||_2^2+\lambda||\bm{\xi}||_1, \label{l1normregularization}
\\ 
&&
Z_1(\mu,\kappa|\V{y},\V{A})=\int\mathrm{d}\bm{\xi}~\mathrm{e}^{-\mu(\mathcal{H}_1(\bm{\xi}|\V{y},\V{A})+\kappa||\bm{\xi}||_0)}, \\
&&
p_1(\bm{\xi};\mu,\kappa|\V{y},\V{A})=\frac{1}{Z_1(\mu,\kappa|\V{y},\V{A})}\mathrm{e}^{-\mu(\mathcal{H}_1(\bm{\xi}|\V{y},\V{A})+\kappa||\bm{\xi}||_0)}.	
	\label{l1canonical}
\end{eqnarray}
The parameter $\mu$ denotes the inverse temperature corresponding to the support estimation by the method of $\ell_1$-norm regularization.
The parameter $\kappa$ is an auxiliary variable introduced to analyze the compression rate $r$ and the limit of $\kappa\to0$ is taken in the end.
The energy function $\mathcal{H}_1$ is exactly the minimized object in (\ref{l1regdef}). We also introduce the averaged free-energy density, given by
\begin{eqnarray}
	f_1(\mu,\kappa)=-\frac{1}{M\mu} \left[\ln Z_1(\mu,\kappa|\V{y},\V{A})\right]_{ \bm{y},\bm{A} },
\end{eqnarray}
which plays the role of the cumulant-generating function that is given by $\phi_0$ in the $\ell_0$ case. In the limit $\mu \to \infty$, the minimizer of the energy function becomes dominant in $p_1$, and we focus on this limit. Any quantity of interest can be calculated from $f_1$. For example, the compression rate $r$ and the distortion $\epsilon_1$ are calculated as 
\begin{eqnarray}
	r&=& \lim_{\mu \to \infty} \lim_{\kappa\to0}\frac{\partial}{\partial \kappa}f_1(\mu,\kappa)
	\Leq{r-derivation}
	,\\
	\epsilon_1&=&
	\lim_{\mu \to \infty}
	\left(1+\mu\frac{\partial}{\partial\mu}-\lambda\frac{\partial}{\partial\lambda}\right)f_1(\mu,0).
	\Leq{epsilon_1-derivation}
\end{eqnarray}

An analytically compact form of $f_1$ is assessed by using the replica method in the limit $M\to \infty$, through the replica identity, as
\be
f_1(\mu,\kappa)=-\lim_{n\to 0}\frac{1}{M\mu n}\ln \lsb Z_1^n(\mu,\kappa|\V{y},\V{A}) \rsb_{\V{y},\V{A}}.
\Leq{replica_l1}
\ee
As in the $\ell_0$ case, we 
assume the replica-symmetric solution. The details of the necessary calculations are presented in Appendix B. The result is given by
\begin{eqnarray}
&&
f_1(\mu \to \infty,\kappa)\nonumber\\
&&=\mathop{\rm extr}\limits_{\hat{\Theta}_1}\left\{\frac{1}{2}\frac{P+\sigma_y^2}{1+\chi_p}-\frac{1}{2}(\hat{P}P-\hat{\chi}_p\chi_p)-\frac{\hat{\chi}_p}{2\alpha\hat{P}}\left((1+2\theta_+\theta_-)\mathrm{erfc}(\theta_+)-\theta_-\frac{2}{\sqrt{\pi}}\mathrm{e}^{-\theta_+^2}\right)\right\},
\label{l1preresult}
\end{eqnarray}
where $\hat{\Theta}_1=\{P,\chi_p,\hat{P},\hat{\chi}_p\}$, $\theta_{\pm}=\frac{\lambda\pm\sqrt{2\kappa\hat{P}}}{\sqrt{2\hat{\chi}_p}}$, and $\mathrm{erfc}(\cdot)$ is the complementary error function, defined as $\mathrm{erfc}(x)=\frac{2}{\sqrt{\pi}}\int_x^\infty\mathrm{d}t\mathrm{e}^{-t^2}$.  The extremization condition gives the following EOSs for the present case:
\subbe
\Leq{EOSs_l1}
\be
&&
\hat{\chi}_p=\frac{P+\sigma_y^2  }{(1+\chi_p)^2},
\Leq{chiphat_l1}
\\ &&
\hat{P}=\frac{ 1 }{1+\chi_p},
\Leq{Phat_l1}
\\ &&
\chi_p=\frac{1}{ \alpha \hat{P} }\left(\mathrm{erfc}(\theta_+)+\sqrt{\frac{\kappa\hat{P}}{\hat{\chi}_p}}\frac{2}{\sqrt{\pi}}\mathrm{e}^{-\theta_+^2}\right), 
\Leq{chip_l1}
\\ &&
P=\frac{\hat{\chi}_p }{ \alpha \hat{P}^2 } \left((1+2\theta_+\theta_-)\mathrm{erfc}(\theta_+)-\theta_-\frac{2}{\sqrt{\pi}}\mathrm{e}^{-\theta_+^2}\right)+\frac{\kappa}{\alpha\hat{P}}\mathrm{erfc}(\theta_+).
\Leq{P_l1}
\ee
\subee
By using \BReqs{r-derivation}{epsilon_1-derivation}, we obtain
\be
&&
r=\frac{1}{\alpha}\mathrm{erfc}(\theta)
\Leq{r_l1}
\\ &&
\epsilon_1=\frac{1}{2}\frac{P+\sigma_y^2}{1+\chi_p}-\frac{1}{2}(\hat{P}P-\hat{\chi}_p\chi_p)-\frac{\hat{\chi}_p}{2\alpha\hat{P}}\left((1-2\theta^2)\mathrm{erfc}(\theta)+\theta\frac{2}{\sqrt{\pi}}\mathrm{e}^{-\theta^2}\right),
\Leq{epsilon_l1}
\ee
where $\theta=\frac{\lambda}{\sqrt{2\hat{\chi}_p}}$.
In addition, a simple formula
\be
\epsilon_1=\frac{1}{2}\hat{\chi}_p.
\Leq{epsilon_l1}
\ee
is derived from the EOSs of \NReq{EOSs_l1} in
the limit of $\kappa\to0$, and a useful relation
\be
\chi_p=\frac{r}{1-r},
\Leq{chip-simples_l1}
\ee
which is similar to \BReq{chi-simples}, 
is offered by \BReqs{EOSs_l1}{r_l1}.
\paragraph{Derivation of $\epsilon^{\rm LS}_1$}
We also evaluate $\epsilon^{\rm LS}_1$, as defined in (\ref{l1pidef}). The computations are rather technical, and there we defer the details to Appendix B. Here, we present an outline of the analysis, and the result.

Again, we use the energy function defined in the $\ell_0$ case, but here the argument is $|\V{\xi}|_0$, determined by $p_1(\V{\xi})$. Thus, we obtain
\begin{eqnarray}
\mathcal{H}_0( |\bm{\xi }|_0 ;\beta|\V{y},\V{A})&=&-\frac{1}{\beta}\ln
\int\mathrm{d}_{ |\bm{ \xi }|_0 }
\bm{x}\mathrm{e}^{-\frac{\beta}{2}||\bm{y}-\bm{A}(|\bm{ \xi}|_0\circ\bm{x})||_2^2}.
\end{eqnarray}
Since the vector $\V{\xi}$ is drawn from $p_1$, we calculate the average value of $(1/M)\mathcal{H}_0(|\bm{ \xi }|_{0})$ over $p_1$, in addition to the configurational average. Taking the limits of $\mu \to \infty$ and then $\beta \to \infty$ afterward, we obtain the desired distortion $\epsilon^{\rm LS}_1$ as follows:
\begin{eqnarray}
	\epsilon^{\rm LS}_1
	=
	\lim_{\beta\to\infty}\lim_{\mu\to\infty}
	\frac{1}{M}
	\left[
	 \int \mathrm{d} \V{\xi }~p_1(\V{\xi};\mu,0|\V{y},\V{A} )
	 \mathcal{H}_0(|\bm{\xi}|_0;\beta|\V{y},\V{A})
	\right]_{\bm{y},\bm{A}}.
	\label{errorafterpi}
\end{eqnarray}
By utilizing the replica method again, we can calculate this. We defer the details of the calculations to Appendix B, and here write down the resultant formula:  
\begin{eqnarray}
	&&\epsilon^{\rm LS}_1
	=\mathop{\rm extr}\limits_{\hat{\Theta}^{\rm LS}_1}
	\Biggl\{\frac{1}{2}\frac{P+\sigma_y^2}{1+\chi_q}\left(\frac{\chi_c}{1+\chi_p}\right)^2-\frac{C+\sigma_y^2}{1+\chi_q}\frac{\chi_c}{1+\chi_p}+\frac{1}{2}\frac{Q+\sigma_y^2}{1+\chi_q}\Biggr.\nonumber\\
	&&-(\hat{C}C-\hat{\chi}_c\chi_c)-\frac{1}{2}(\hat{Q}Q-\hat{\chi}_q\chi_q)
	\nonumber\\
	&&\Biggl.-\frac{\hat{\chi}_p}{2\alpha\hat{Q}}\left(\left(\frac{\hat{\chi}_q}{\hat{\chi}_p}-2\frac{\hat{\chi}_c}{\hat{\chi}_p}\frac{\hat{C}}{\hat{P}}+(1+2\theta^2)\frac{\hat{C}^2}{\hat{P}^2}\right)\mathrm{erfc(\theta)}+\theta\left(\frac{\hat{\chi}_c^2}{\hat{\chi}_p^2}-\frac{\hat{C}^2}{\hat{P}^2}\right)\frac{2}{\sqrt{\pi}}\mathrm{e}^{-\theta^2}\right)\Biggr\},
	\Leq{epsilon_1^PI}
\end{eqnarray}
where $\hat{\Theta}^{\rm LS}_1=\{C,\chi_c,Q,\chi_q,\hat{C},\hat{\chi}_c,\hat{Q},\hat{\chi}_q\}$, and $\theta=\frac{\lambda}{\sqrt{2\hat{\chi}_p}}$. One point to remark on is that we should not take the extremization condition with respect to $\hat{\Theta}_1=\{P,\chi_p,\hat{P},\hat{\chi}_p\}$ in this expression. Instead, we should substitute the extremizer of (\ref{l1preresult}) into it. Applying the extremization condition with respect to $\hat{\Theta}^{\rm LS}$ gives 
\subbe
\Leq{EOSs_l1+PI}
\be
&&
\hat{\chi}_q=\frac{P+\sigma_y^2}{(1+\chi_q)^2}\left(\frac{\chi_c}{1+\chi_p}\right)^2-2\frac{C+\sigma_y^2}{(1+\chi_q)^2}\frac{\chi_c}{1+\chi_p}+\frac{Q+\sigma_y^2}{(1+\chi_q)^2},
\Leq{chiqhat_l1+PI}
\\ &&
\hat{Q}=\frac{1}{1+\chi_q},
\Leq{Qhat_l1+PI}
\\ &&
\hat{\chi}_c=-\frac{P+\sigma_y^2}{1+\chi_q}\frac{\chi_c}{(1+\chi_p)^2}+\frac{C+\sigma_y^2}{1+\chi_q}\frac{1}{1+\chi_p}
\Leq{chihatc_l1+PI}
\\ &&
\hat{C}=-\frac{1}{1+\chi_q}\frac{\chi_c}{1+\chi_p} ,
\Leq{Chat_l1+PI}
\\ &&
\chi_q=\frac{1}{\alpha\hat{Q}}\mathrm{erfc}(\theta),
\Leq{chiq_l1+PI}
\\ &&
Q=\frac{\hat{\chi}_p}{\alpha\hat{Q}^2}\left(\left(\frac{\hat{\chi}_q}{\hat{\chi}_p}-2\frac{\hat{\chi}_c}{\hat{\chi}_p}\frac{\hat{C}}{\hat{P}}+(1+2\theta^2)\frac{\hat{C}^2}{\hat{P}^2}\right)\mathrm{erfc(\theta)}+\theta\left(\frac{\hat{\chi}_c^2}{\hat{\chi}_p^2}-\frac{\hat{C}^2}{\hat{P}^2}\right)\frac{2}{\sqrt{\pi}}\mathrm{e}^{-\theta^2}\right),
\Leq{Q_l1+PI}
\\ &&
\chi_c=\frac{1}{\alpha\hat{Q}}\left(-\frac{\hat{C}}{\hat{P}}\mathrm{erfc}(\theta)+\theta\frac{\hat{\chi}_c}{\hat{\chi}_p}\frac{2}{\sqrt{\pi}}\mathrm{e}^{-\theta^2}\right),
\Leq{chic_l1+PI}
\\ &&
C=-\frac{\hat{\chi}_p}{\alpha\hat{Q}}\left(\left(-\frac{\hat{\chi}_c}{\hat{\chi}_p}\frac{1}{\hat{P}}+(1+2\theta^2)\frac{\hat{C}}{\hat{P}^2}\right)\mathrm{erfc}(\theta)-\theta\frac{\hat{C}}{\hat{P}^2}\frac{2}{\sqrt{\pi}}\mathrm{e}^{-\theta^2}\right)
\Leq{C_l1+PI}.
\ee
\subee
From the EOSs, we can obtain the following simple relations:
\subbe
\be
&&
\chi_q=\frac{r}{1-r}=\chi_p,
\\ &&
\hat{Q}=\hat{P}=\frac{1}{1+\chi_p},
\\ &&
\epsilon_{1}^{\rm LS}=\frac{1}{2}\hat{\chi}_q.
\Leq{epsilon_1^PI-simples}
\ee 
\Leq{simples_l1+PI}
\subee

We now make some comments regarding the derivation of \BReq{epsilon_1^PI}. In order to calculate the configurational average, we are required to deal with two different factors, $Z_1$ in $p_1=(1/Z_1)\mathrm{e}^{-\mu \mathcal{H}_1}$, and the logarithm in $\mathcal{H}_0$. Correspondingly, as in the $\ell_0$ case, we introduce replicas of two different kinds: $n$ replicas to handle $1/Z_1$, and $\nu$ replicas to handle the logarithm. Using them, we can rewrite (\ref{errorafterpi}) as
\be
\epsilon^{\rm LS}_1=
\lim_{\beta\to\infty}\lim_{\mu \to \infty}
\lim_{n\to 0}\lim_{\nu \to 0} 
-\frac{1}{M\beta\nu}
\ln
\left[
Z^{n-1}_1(\mu,0|\bm{y},\bm{A})
\int\mathrm{d}\bm{\xi} 
 \mathrm{e}^{-\mu\mathcal{H}_1(\bm{\xi}|\bm{y},\bm{A})}
\lb 
\int\mathrm{d}_{|\V{\xi}|_0}\bm{x}
\mathrm{e}^{-\frac{\beta}{2}||\bm{y}-\bm{A}(|\bm{\xi}|_0\circ\bm{x})||_2^2}
\rb^\nu
\right]_{\bm{y},\bm{A}}.
\Leq{replica_l1+PI}
\end{eqnarray}
It is now possible to calculate the configurational average by assuming $n$ and $\nu$ are natural numbers, and we can follow the usual prescription of the replica method. However, there remains a technical point concerning the limits $n\to 0$ and $\nu \to 0$ in the present formulation. The region around $n=\nu=0$ has an unusual property. The extremization condition with respect to the order parameters yields several different solutions. Among these solutions, by employing a versatile tool of spin-glass theory to analyze a probabilistic model conditioned by another probabilistic model, called the Franz-Parisi potential, we should choose the one analytically connected to $\hat{\Theta}_{1}$ in (\ref{l1preresult}) in the limit $\nu \to 0$. This is achieved by the remark given below \BReq{epsilon_1^PI}~\cite{Franz97}. 

\subsection{Numerical validation using simulations on finite $M$}

\subsubsection{$\ell_0$-based method}
We examine the analytical results, using numerical simulations of finite-size systems.
When $M$ is sufficiently small, we can obtain the cumulant-generating function $\phi_0$ by exhaustively searching all possible combinations of basis vectors.
In cases where $M$ is less small, we use the exchange Monte Carlo (MC) method to sample basis vector combinations obeying the canonical distribution at various temperature points \cite{Swendsen86,Hukushima96}, and then estimate the cumulant-generating function $\phi_0$ using the multi-histogram method \cite{Ferrenberg88}.

In all simulations, we set $\alpha=0.5$ and $\sigma_y^2=1$.
We treat two values of $r$ equal to $0.2$ and $0.4$.
In the case of $r=0.2~(0.4)$, we calculate cumulant-generating function values at 15 temperature points, which are distributed according to the geometric progression in the range between 1 and 10 (between 1 and 35) in the value of $\mu$.
We conduct the exhaustive search for $M\leq 25~(15)$, and use the exchange MC method for larger $M$.
The configurational average is calculated by taking the median over 1000 different samples of $(\bm{y},\bm{A})$.
The error bars are estimated by the Bootstrap method.

The procedure for our MC method will now be explained.
At every temperature point, we randomly choose the initial vector $\bm{c}$ among those satisfying $||\bm{c}||_0=rM$.
For $r=0.2$, the number of MC steps required for thermalization and sufficient sampling is $2,3,4,7,10\times 10^4$ for $M=30,35,40,45,50$, respectively, while for $r=0.4$ it is $2,4,8,15,30\times 10^4$ for $M=20,25,30,35,40$, respectively.
The first half of the MC steps are discarded for thermalization.  
One MC step consists of two parts. First, updating once at every temperature point, and then exchanging once between every pair of neighboring temperature points. 
In each update of $\bm{c}$, we randomly choose one index $i$ such that $c_i=1$ and another $j$ such that $c_j=0$ to flip into the opposite state. That is, we set $c_i=0$ and $c_j=1$, and accept or reject this trial according to the Metropolis criterion based on the energy values calculated from $\mathcal{H}_0$ (\ref{l0energy}). The Metropolis criterion is also used in the exchange of $\bm{c}$s of different temperature points.

\begin{figure}[t]
\begin{tabular}{ccc}
\begin{minipage}{0.33\hsize}
\begin{center}
\includegraphics[width=\hsize]{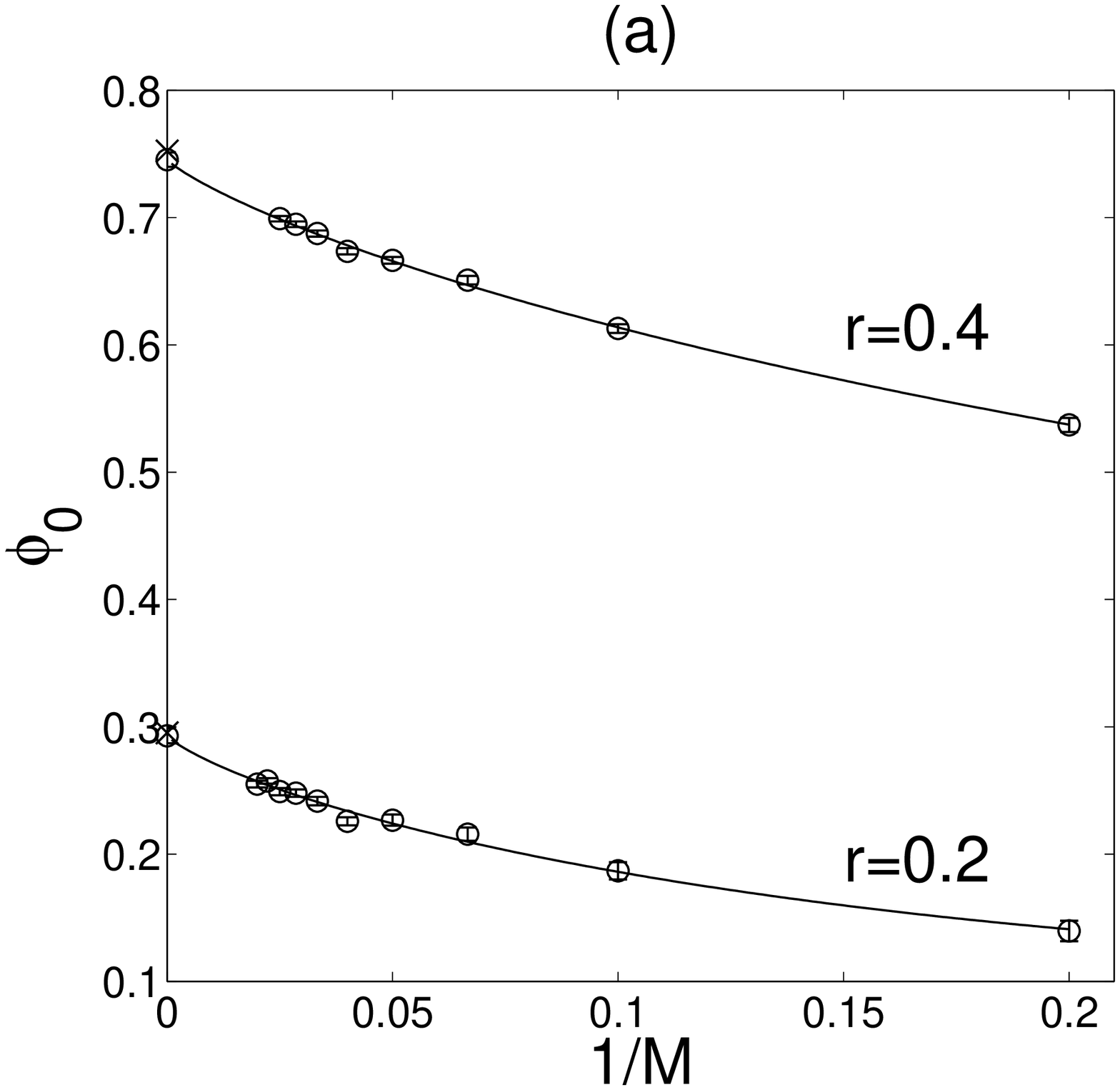}
\end{center}
\end{minipage}
\begin{minipage}{0.33\hsize}
\begin{center}
\includegraphics[width=\hsize]{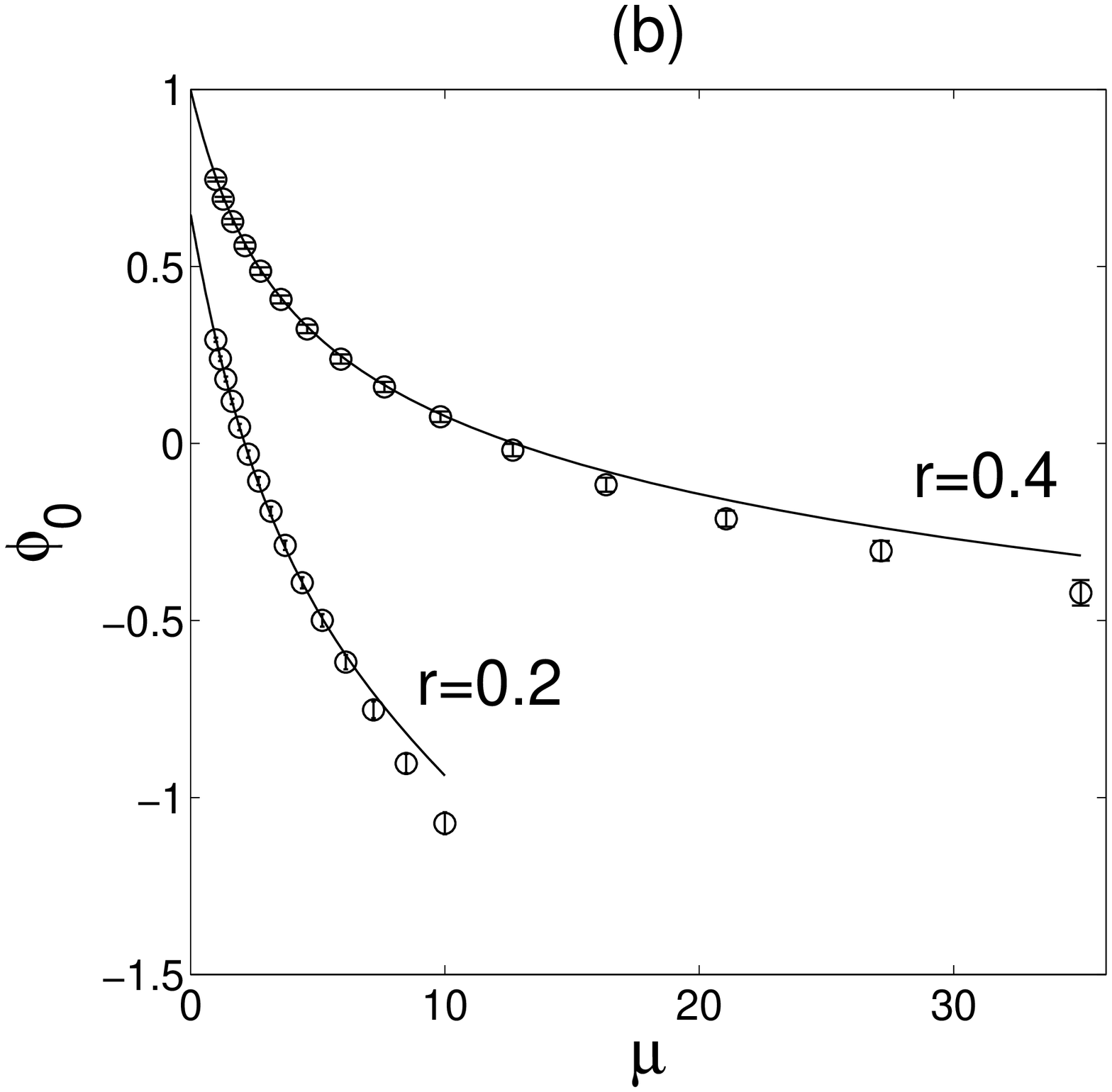}
\end{center}
\end{minipage}
\begin{minipage}{0.33\hsize}
\begin{center}
\includegraphics[width=\hsize]{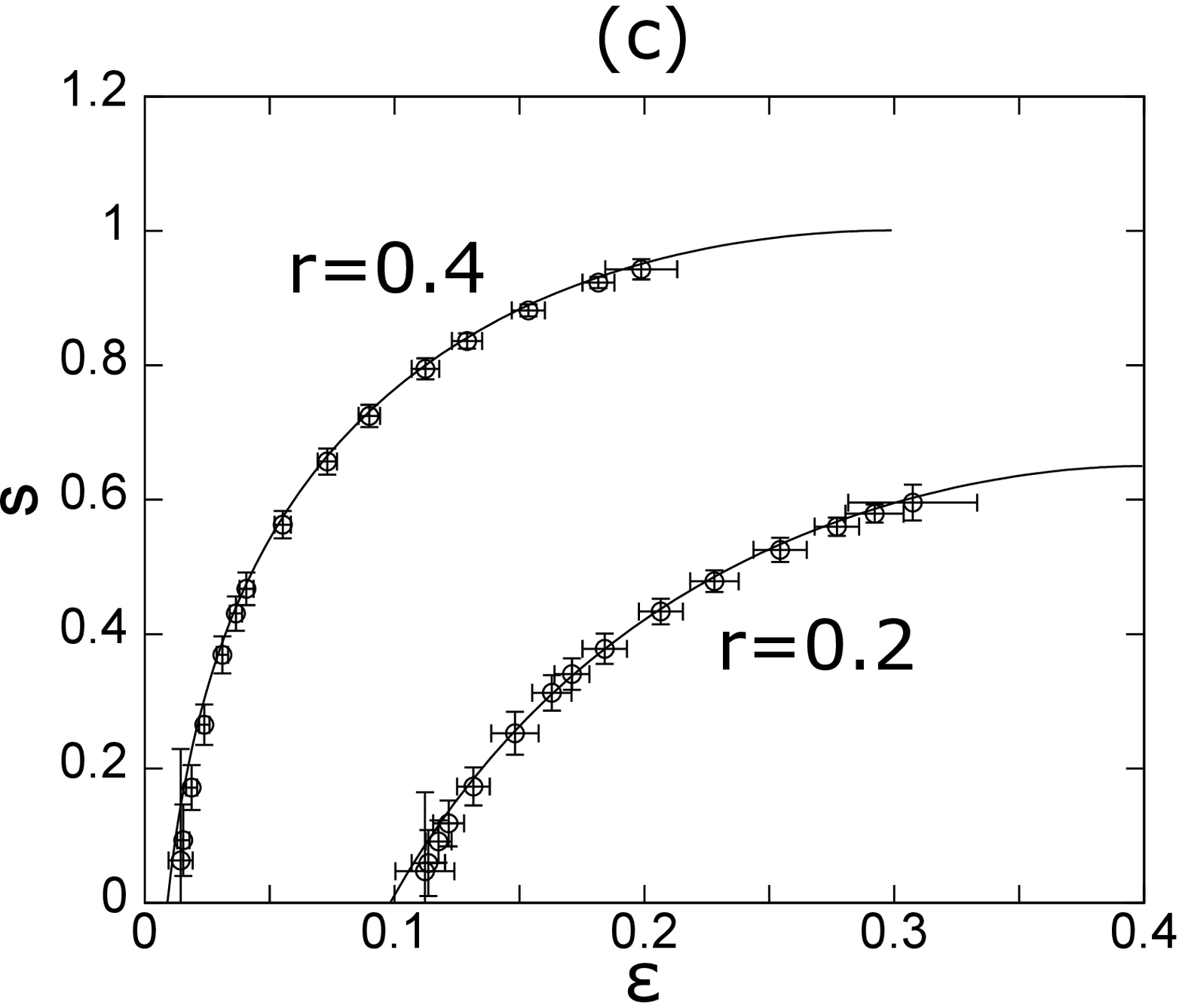}
\end{center}
\end{minipage}
\end{tabular}
\caption{Cumulant-generating function $\phi_0$ and entropy density $s$ of the $\ell_0$-based method with $\sigma_y^2=1$, $\alpha=0.5$, and $r=0.2,0.4$.
(a) Plots of numerically evaluated $\phi_0$ at $\mu=1$.
The lines are given by the linear regression.
On the vertical axis, the circles and crosses represent the extrapolated and analytical values in the $M\to\infty$ limit, respectively.
The lengths of the error bars are comparable to the sizes of symbols.
(b) Plots of $\phi_0$ in the $M\to\infty$ limit.
The lines and circles represent the analytical and extrapolated values, respectively.
The lengths of the error bars are comparable to the sizes of symbols.
(c) Plots of $s$ against $\epsilon$ in the $M\to\infty$ limit.
The lines and circles represent the analytical and extrapolated values, respectively.
These are calculated from the values of $\phi_0$ in (b).}
\label{nvl0}
\end{figure}

The results of the numerical simulations are presented in figure \ref{nvl0}.
Figure \ref{nvl0}(a) shows the results of the cumulant-generating function value at $\mu=1$.
On the vertical axis, the circles represent extrapolated values from finite-size results.
The extrapolation lines are given by the linear regression using an asymptotic form $\phi_0\approx a+bM^{-1}+cM^{-1}\ln M^{-1}$.
The regression is conducted by employing the method of least squares, as follows:
\begin{eqnarray}
	\min_{a,b,c}\frac{1}{2}\sum_M\left(a+b\frac{1}{M}+c\frac{1}{M}\ln\frac{1}{M}-\phi_0(M)\right)^2.\label{extrap}
\end{eqnarray}
The asymptotic form is based on the Stirling's formula and is exact at $\mu=0$, which motivates us to use the form even for $\mu\not=0$.
The cumulant-generating function and entropy density in the limit $M\to\infty$ are presented in figures \ref{nvl0}~(b) and \ref{nvl0}~(c), respectively.
The lines represent the analytical results.
The circles represent the extrapolated values from the numerical results.
The analytical solutions are seen to be consistent with the numerical ones.
Hence, the numerical results clearly validate the analytical results in the $\ell_0$-based method.

\subsubsection{$\ell_1$-based method}
Similarly to the case of the $\ell_0$-based method, we examine the analytical results of the $\ell_1$-based method by performing numerical simulations on finite-size systems.
We carry out the $\ell_1$-norm regularization using quadratic programming, and evaluate the distortion before the method of LS, $\epsilon_1$; the distortion after the method of LS, $\epsilon^{\rm LS}_1$; and the compression rate $r$.

The values of $\alpha$ and $\sigma_y^2$ are fixed as $\alpha=0.5$ and $\sigma_y^2=1$ for all simulations.
We treat two values of $\lambda$ equal to $1$ and $2$.
We calculate (\ref{l1regdef}) and (\ref{l1pidef}) using quadratic programming and the method of LS for $M=50,100,\dots,250$.

\begin{figure}[t]
\begin{tabular}{ccc}
\begin{minipage}{0.33\hsize}
\begin{center}
\includegraphics[width=\hsize]{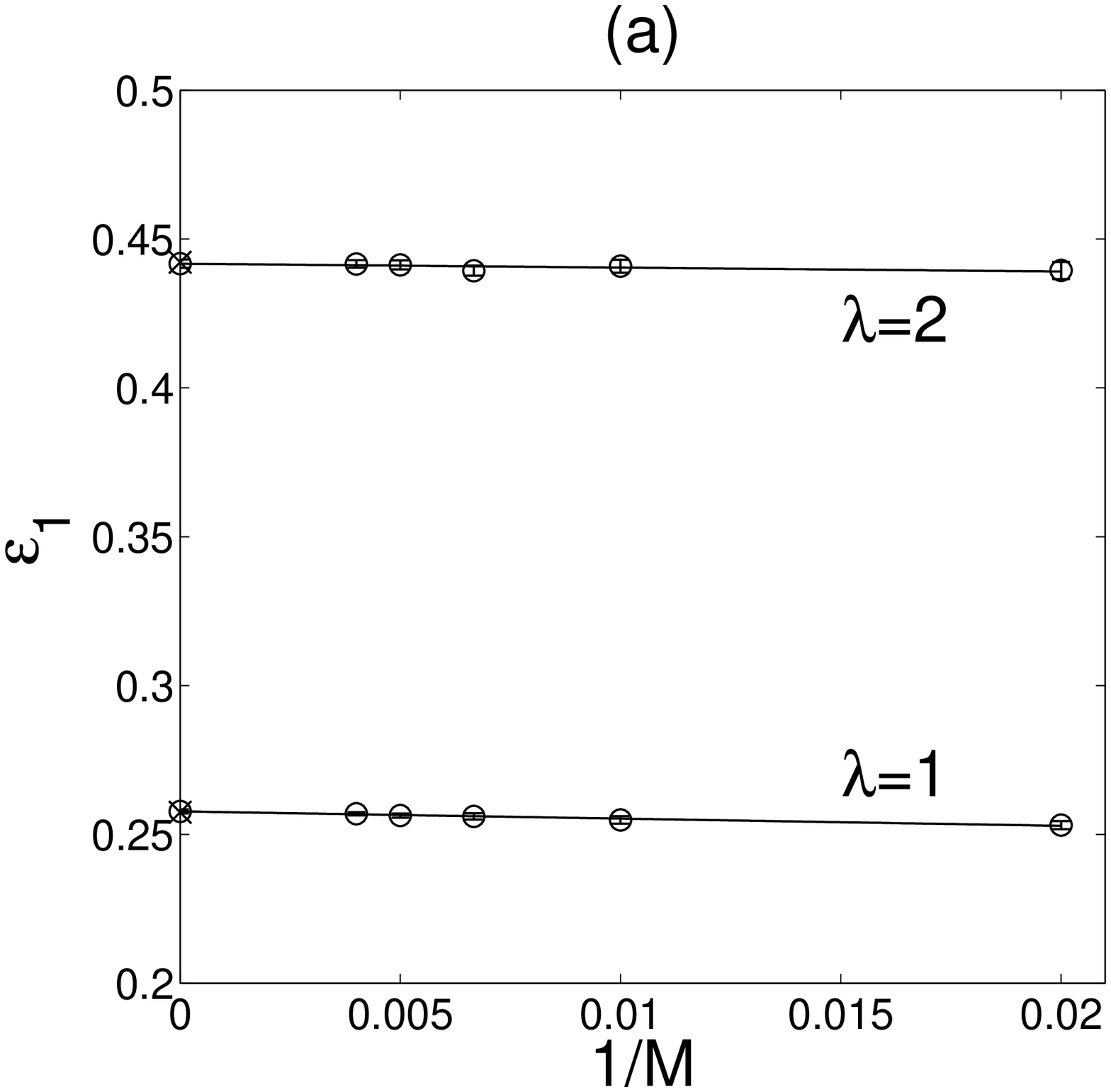}
\end{center}
\end{minipage}
\begin{minipage}{0.33\hsize}
\begin{center}
\includegraphics[width=\hsize]{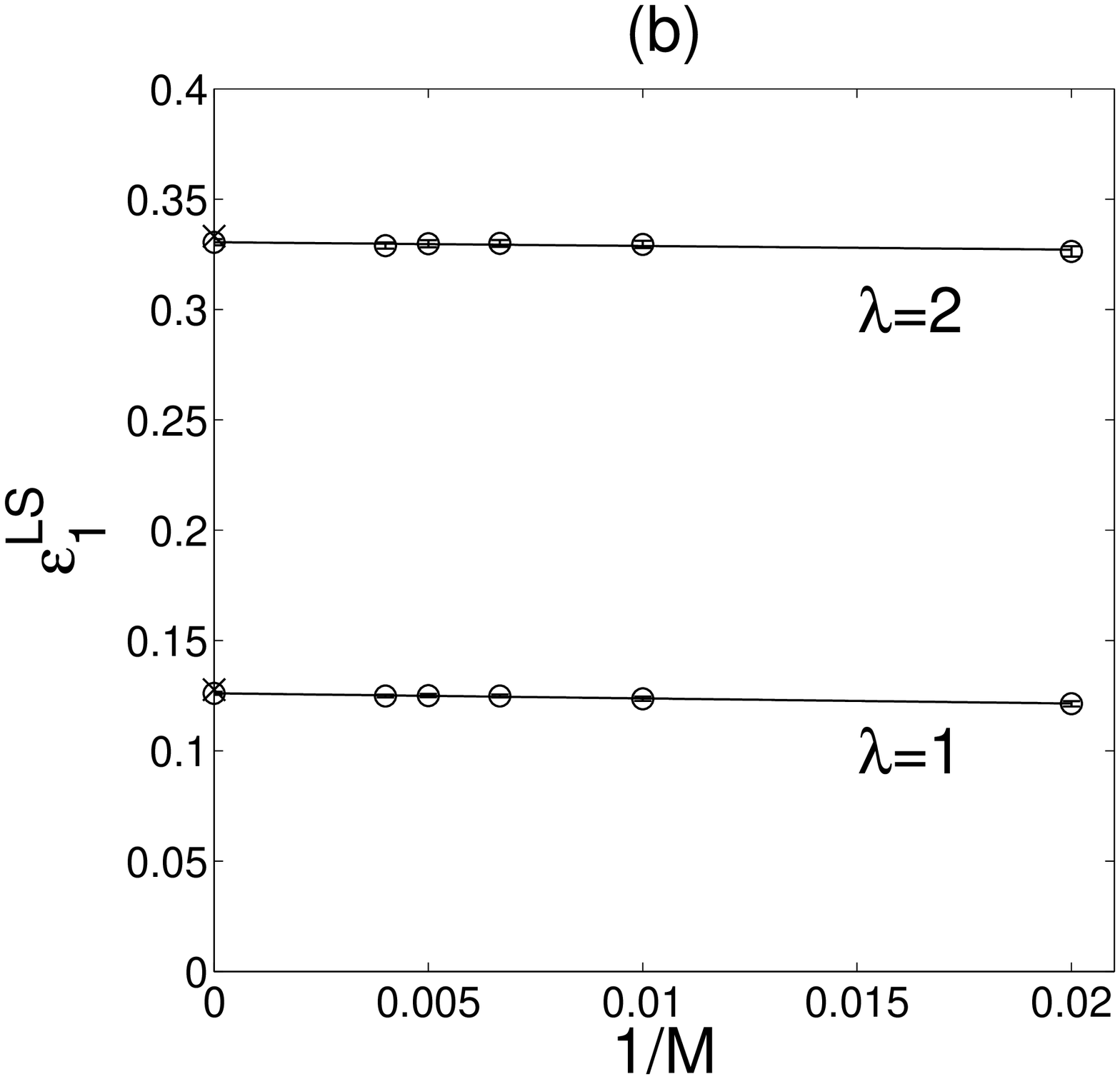}
\end{center}
\end{minipage}
\begin{minipage}{0.33\hsize}
\begin{center}
\includegraphics[width=\hsize]{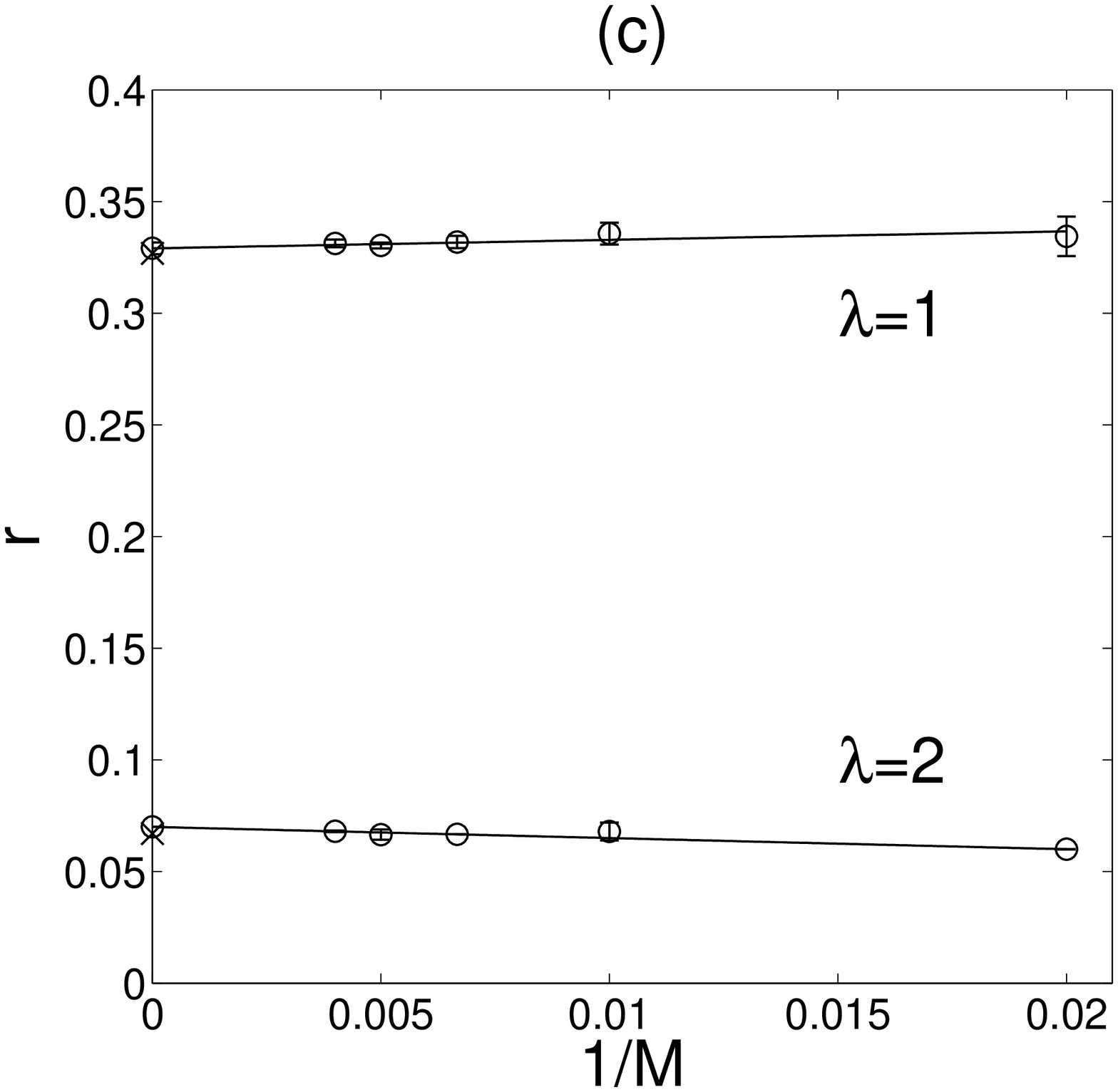}
\end{center}
\end{minipage}
\end{tabular}
\caption{Plots of numerically evaluated values of the $\ell_1$-based method with $\sigma_y^2=1$, $\alpha=0.5$, and $\lambda=1,2$.
The extrapolation lines are given by the linear regression.
On the vertical axes, the circles and crosses represent the extrapolated and analytical values in the $M\to\infty$ limit, respectively.
The lengths of the error bars are comparable to the sizes of symbols.
(a) Distortion before the method of LS, $\epsilon_1$.
(b) Distortion after the method of LS, $\epsilon_1^{\rm LS}$.
(c) Compression rate, $r$.}
\label{nvl1}
\end{figure}

The results of the numerical simulations are shown in figure \ref{nvl1}.
Figures \ref{nvl1}(a)--(c) plot the numerically evaluated distortion before the method of LS, distortion after the method of LS, and the compression rate, respectively, against the system size $M$.  
On the vertical axes, the circles and crosses represent extrapolated and analytical values in the $M\to\infty$ limit, respectively.
The extrapolation lines are given by the linear regression using 
the asymptotic forms $\epsilon_1\approx a+bM^{-1}$, $\epsilon_1^{\rm LS}\approx c+dM^{-1}$, and $r\approx e+fM^{-1}$.
We see that the analytical solutions are very close to the extrapolated values.
This correlation clearly demonstrates the reliability of the analytical results.

\subsection{Comparison in the trade-off relation}
\begin{figure}[t]
\begin{tabular}{cc}
\begin{minipage}{0.5\hsize}
\begin{center}
\includegraphics[width=\hsize]{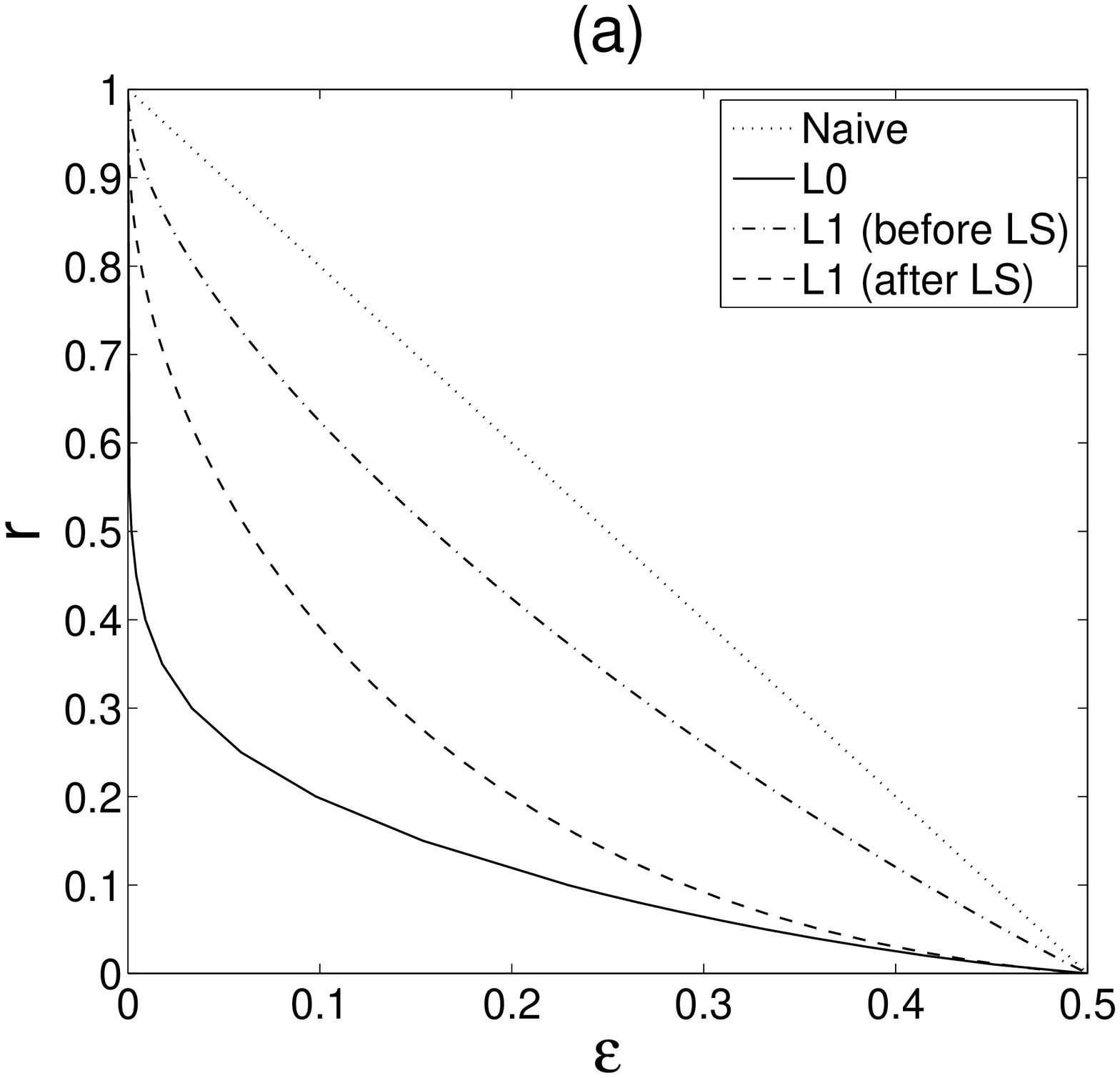}
\end{center}
\end{minipage}
\begin{minipage}{0.5\hsize}
\begin{center}
\includegraphics[width=\hsize]{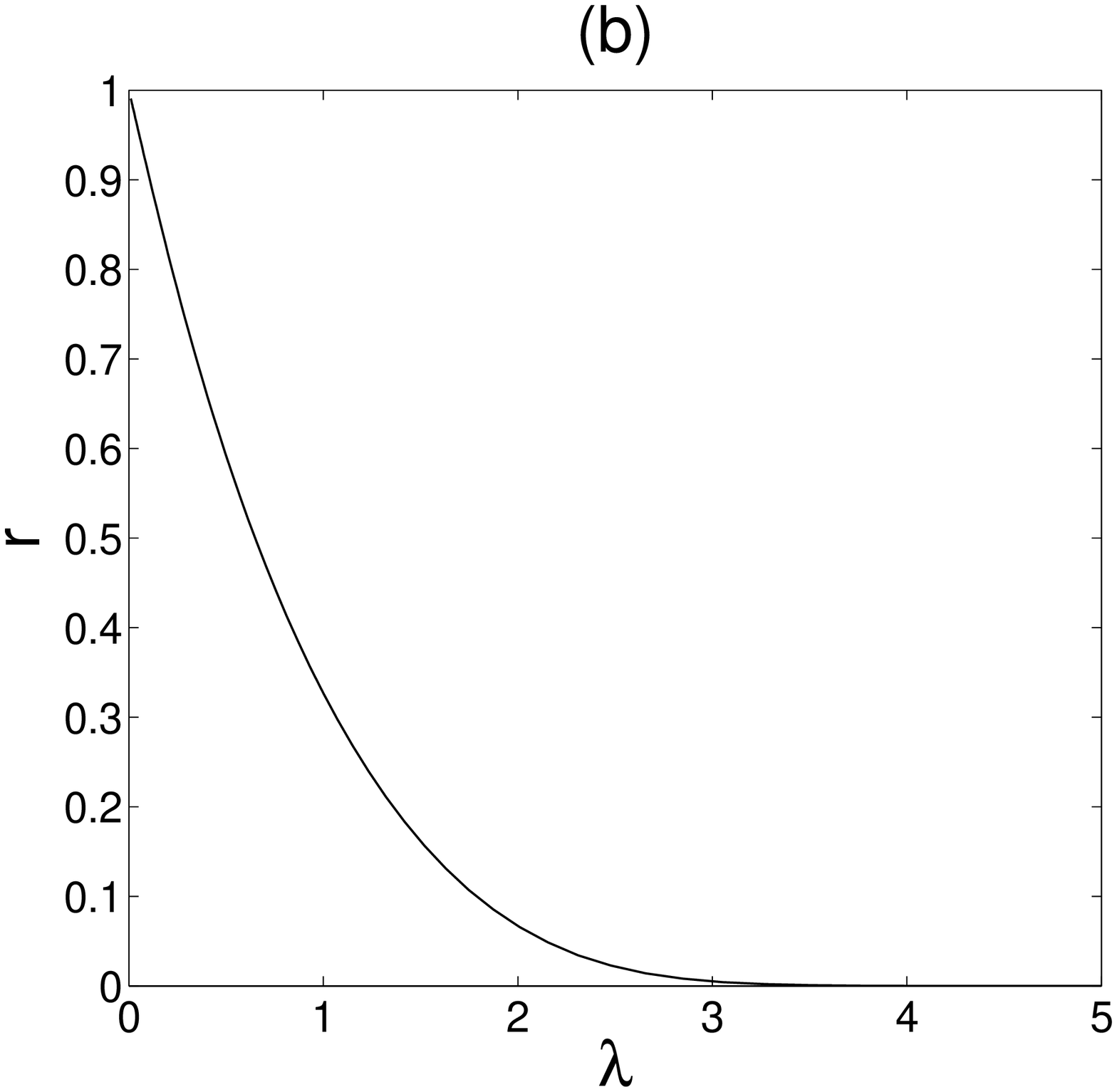}
\end{center}
\end{minipage}
\end{tabular}
\caption{Results of the analysis in the $M\to\infty$ limit with $\sigma_y^2=1$ and $\alpha=0.5$. (a) Trade-off relations of the naive, $\ell_0$-based, and $\ell_1$-based methods (before and after the method of LS). (b) Relation between the rate and the regularization coefficient in the $\ell_1$-based method.}
\label{RD}
\end{figure}

We compare the ideal performance in the $M\to\infty$ limit for different methods in terms of the trade-off relation between the representation distortion and the compression rate.
Figure \ref{RD}(a) shows the trade-off relations in the case of $\alpha=0.5$.
We see that both of the OCB-based methods achieve a better trade-off relation than the naive one.
In the OCB-based strategy, the $\ell_0$-based method significantly outperforms the $\ell_1$-based one, even if the method of LS is operated after carrying out support estimation by the $\ell_1$-norm regularization.
We attribute the inferiority of the $\ell_1$-based method to the regularization term.
Indeed, as shown in figure \ref{RD}~(b), the regularization term is necessary to decrease the rate, 
but it distorts the original purpose of minimizing the distortion, as clearly seen from (\ref{l1normregularization}).

\begin{figure}[t]
\begin{tabular}{cc}
\begin{minipage}{0.5\hsize}
\begin{center}
\includegraphics[width=\hsize]{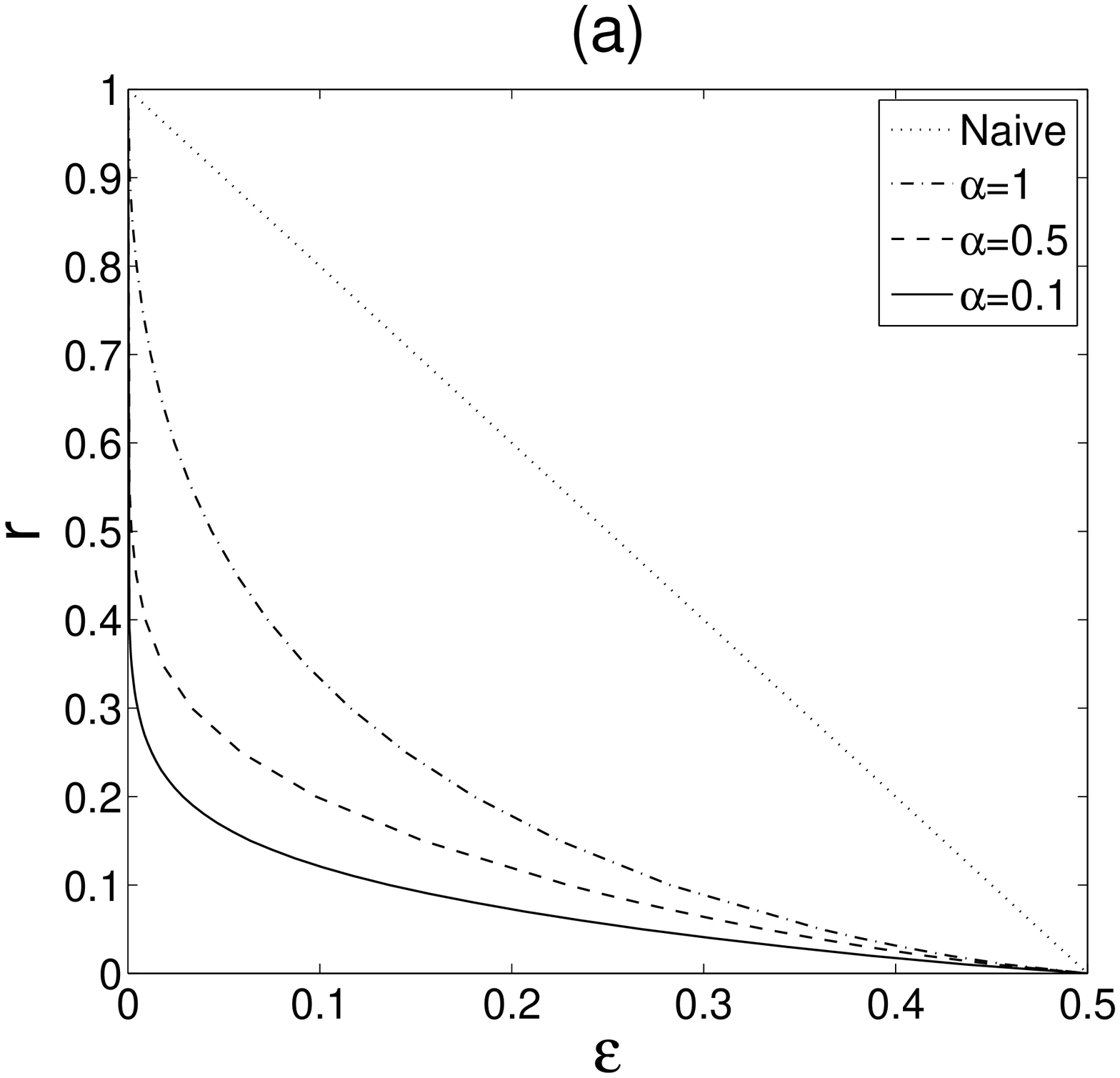}
\end{center}
\end{minipage}
\begin{minipage}{0.5\hsize}
\begin{center}
\includegraphics[width=\hsize]{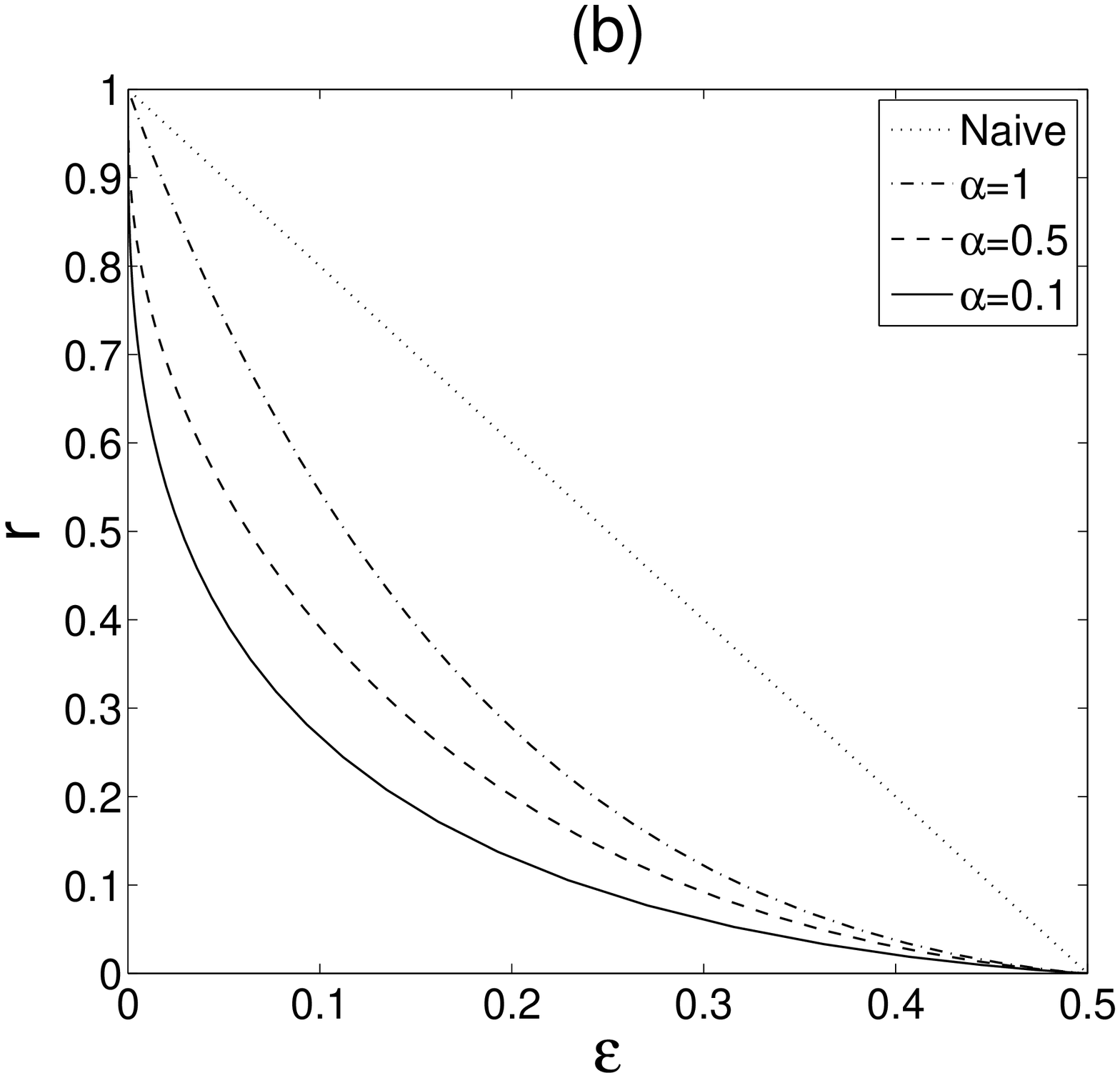}
\end{center}
\end{minipage}
\end{tabular}
\caption{Trade-off relations at various values of $\alpha$ in the case of $\sigma_y^2=1$. (a) Results of the $\ell_0$-based method. (b) Results of the $\ell_1$-based method (after the method of LS). For both the methods, against fixed a $r$, the distortion $\epsilon$ becomes smaller as $\alpha$ decreases.}
\label{increased}
\end{figure}

For a further comparison of the OCB-based methods, figure \ref{increased} shows the trade-off relations where different values of $\alpha$ control the degree of overcompleteness.
figures \ref{increased}~(a) and \ref{increased}~(b) present the results of the $\ell_0$- and $\ell_1$-based methods, respectively.
In the $\ell_1$-based method, the method of LS has been operated after the support estimation. 
Both methods achieve a better trade-off relation as the degree of overcompleteness increases, or $\alpha$ decreases.
Another interesting observation is the superiority of the $\ell_0$-based method compared to the $\ell_1$-based one, regardless of the degree of overcompleteness.

\subsubsection{In the large limit of the degree of overcompleteness, $\alpha \to 0$ }
From figure \ref{increased} we see that the distortion becomes smaller as $\alpha$ decreases, both for the $\ell_0$- and $\ell_1$-based methods. An interesting question is whether the distortion vanishes or not in the limit $\alpha \to 0$, or more quantitatively, how $\epsilon$ is scaled by $\alpha$ in the small limit. 

Deferring the detailed calculations to \Rapp{alphazero_l0} and \ref{sec:alphazero_l1}, here we summarize our analytical results on the behavior of $\epsilon$ in the limit $\alpha \to 0$
\be
&&
\epsilon_{0}\propto \alpha^{\frac{2r}{1-r}} \to 0,
\Leq{epsilon_0-alphazero}
\\&&
\epsilon_{1}\to \frac{1}{2} (1-r)^2 \sigma_y^2=O(1),
\Leq{epsilon_1-alphazero}
\\&&
\epsilon^{\rm LS}_{1} \propto |\ln \alpha|^{-1} \to 0.
\Leq{epsilon_1^PI-alphazero}
\ee
The asymptotic behaviors of $\epsilon_0$ and $\epsilon_1^{\rm LS}$ are examined using numerical solutions of the corresponding EOSs, \BReqs{EOSs_l0}{EOSs_l1+PI}, in \Rfig{alpha-epsilon}. Our analytic formulas show an excellent agreement with the numerical results. 
\begin{figure}[t]
\begin{center}
\includegraphics[width=0.45\hsize]{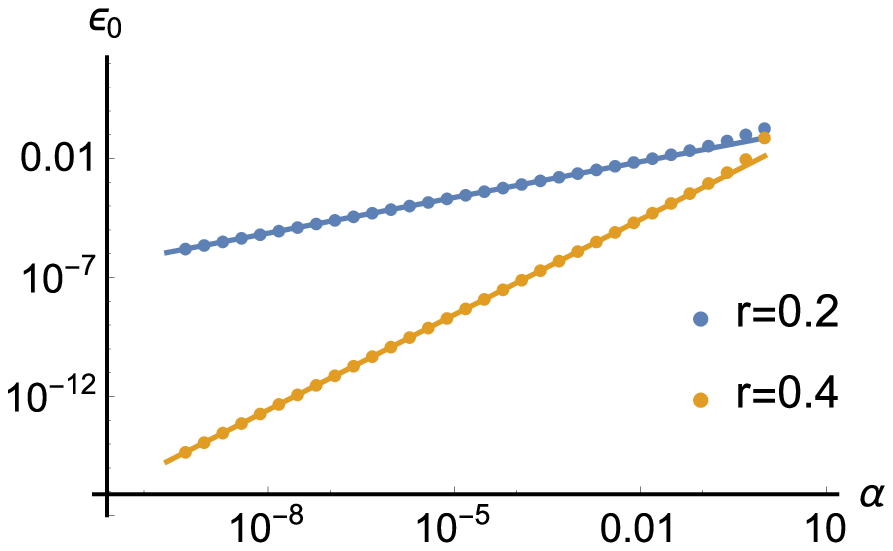}
\includegraphics[width=0.45\hsize]{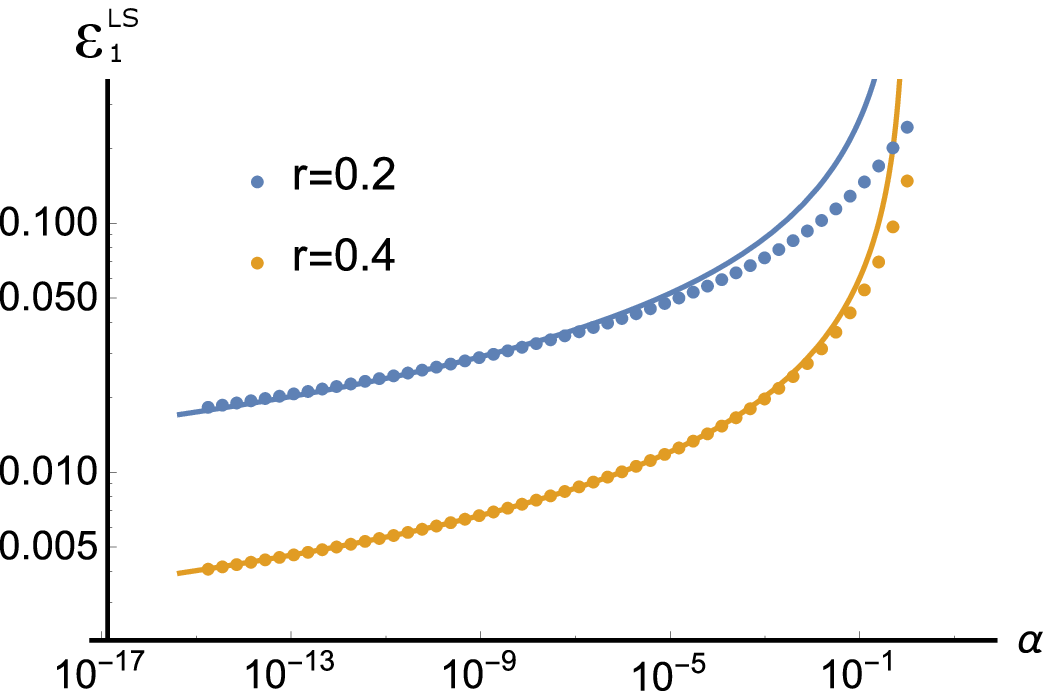}
\end{center}
\caption{
Plots of $\epsilon$ against $\alpha$ in the small $\alpha$ limit, derived by solving \BReqs{EOSs_l0}{EOSs_l1+PI} numerically. The left panel represents $\epsilon_0$, and the right panel represents $\epsilon_{1}^{\rm LS}$. The lines are the fits based on our analytical formulas, \BReqs{epsilon_0-alphazero}{epsilon_1^PI-alphazero}, and these show excellent agreement with the points obtained by the numerical evaluations. }
\Lfig{alpha-epsilon}
\end{figure}

We stress the consequence of \BReqss{epsilon_0-alphazero}{epsilon_1^PI-alphazero}. First, they give a firm indication that it is reasonable to apply the method of LS after the $\ell_1$-norm regularization, which is heuristically employed in related problems such as compressed sensing in practical situations. The difference in \BReqs{epsilon_1-alphazero}{epsilon_1^PI-alphazero} indicates that the method of LS actually diminishes the distortion, and even eliminates it in the ideal limit $\alpha \to 0$, which never happens with only the use of $\ell_1$-norm regularization. Second, \BReq{epsilon_0-alphazero} provides a general bound for the computational cost of searching the appropriate basis vectors. From \BReq{epsilon_0-alphazero}, given a target value of the distortion $\hat{\epsilon}$ and some data on the length $M$, the required size $N_{\rm req}(\hat{\epsilon},M)$ of the basis matrix to achieve this distortion value is scaled as 
\be
N_{\rm req}(\hat{\epsilon},M)\propto  M \hat{\epsilon}^{-\frac{1-r}{2r}}.~~ (\ell_0)
\ee
This grows in a polynomial manner as the target distortion value $\hat{\epsilon}$ decreases, and the exponent of the polynomial negatively grows as the compression rate $r$ decreases. This quantitative information will provide a theoretical basis in designing algorithms. Finally, \BReq{epsilon_1^PI-alphazero} manifests the limit of the $\ell_1$-based method. The size $N_{\rm req}$ required to achieve the target distortion $\hat{\epsilon}$ in this case is scaled as 
\be
N_{\rm req}(\hat{\epsilon},M)\propto  M \mathrm{e}^{\frac{1}{\hat{\epsilon}}},~~(\ell_1+{\rm LS}).
\Leq{N_req_l1+PI}
\ee
This grows exponentially as $\hat{\epsilon}$ decreases, which is considered to be reasonable. If it were a polynomial, versatile algorithms exactly solving the $\ell_1$-norm regularization could be applied to solve the problem with a computational cost of a polynomial order of the system size and the precision, which is believed not to be possible. However, \BReq{N_req_l1+PI} can still be useful, because it provides a quantitative comparison between the data size $M$ and the acceptable distortion $\hat{\epsilon}$ in an unified manner.

\section{Examination of practical performance}\label{practice}

\subsection{Algorithms and their performances}
A lot of computational time is required to conduct the exhaustive search used in the $\ell_0$-based method.
However, it is considered that certain greedy algorithms might work well for practical applications.
Orthogonal matching pursuit (OMP, figure \ref{OMP}) is a greedy algorithm that may be suitable for the present purpose~\cite{Pati93,Davis94}.
\begin{figure}[t]
\begin{center}
\begin{tabular*}{\linewidth}{l}\hline
{\it Input:} a data vector $\bm{y}$, a basis matrix $\bm{A}$, a rate $r$.\\
{\it Initialization:} $\bm{x}^{(0)}=\bm{0}$, $U=\{1,2,\ldots,N\}$, $S^{(0)}=\emptyset$.\\
{\it Iteration:} repeat from $n=1$ until $n=rM$:\\
\ \ \ \ \ $\bm{r}=\bm{y}-\bm{Ax}^{(n-1)}$,\\
\ \ \ \ \ $j=\mathop{\rm arg~max}
\limits_{k\in U \backslash S^{(n-1)}} \{|\bm{a}_k^{\rm T}\bm{r}|\}$,\\
\ \ \ \ \ $S^{(n)}=S^{(n-1)}\cup\{j\}$,\\
\ \ \ \ \ $\bm{x}^{(n)}=\mathop{\rm arg~min}\limits_{\bm{x}}\{||\bm{y}-\bm{Ax}||_2\}$\ \ \ subj. to $\mathrm{supp}(\bm{x})\subset S^{(n)}$.\\
{\it Output:} a compressed vector $\hat{\bm{x}}=\bm{x}^{(rM)}$.\\ \hline
\end{tabular*}
\end{center}
\caption{The procedure of OMP. $\emptyset$ is the empty set. 
$\mathrm{supp}(\cdot)$ is the support set.}
\label{OMP}
\end{figure}
OMP only requires a computational time of order $O(M^4)$
for the current purpose.
We compare the performance of OMP with the ideal performances of both the $\ell_0$- and $\ell_1$-based methods.

In addition to OMP, we also examine approximate message passing (AMP), as a representative algorithm carrying out the $\ell_1$-norm regularization.
From the viewpoint of 
quadratic 
programming, $\ell_1$-norm regularization is solved exactly using versatile algorithms, which require a computational time of order $O(M^3)$.
In contrast, AMP only requires a computational time of order $O(M^2)$
per update.
Despite the low computational cost, AMP is known to be able to recover the results of those versatile algorithms, in certain reasonable situations \cite{Donoho09}.
The present case, where the basis matrix $\bm{A}$ and the data vector $\bm{y}$ are generated from i.i.d. normal distributions, is expected to be one such situation.
Hence, we can fairly compare the result of AMP with the ideal performance of the $\ell_1$-based method, and therefore with that of OMP.
\begin{figure}[t]
\begin{center}
\begin{tabular*}{\linewidth}{l}\hline
{\it Input:} a data vector $\bm{y}$, a basis matrix $\bm{A}$, a regularization coefficient $\lambda$, a tuning parameter $\delta$.\\
{\it Initialization:} $\bm{x}^{(0)}=\bm{0}$, $\chi^{(0)}=0$, $\bm{r}^{(0)}=\bm{y}$.\\
{\it Iteration:} repeat until convergence at $n=\hat{n}$:\\
\ \ \ \ \ $\hat{Q}=\frac{1}{1+\chi^{(n-1)}}$,\\
\ \ \ \ \ $\bm{r}^{(n)}=(1-\hat{Q})\bm{r}^{(n-1)}+\hat{Q}(\bm{y}-\bm{Ax}^{(n-1)})$,\\
\ \ \ \ \ $\bm{h}=\bm{A}^{\rm T}\bm{r}^{(n)}+\hat{Q}\bm{x}^{(n-1)}$,\\
\ \ \ \ \ $\chi^{(n)}=(1-\delta)\chi^{(n-1)}+\delta\frac{1}{\hat{Q}}\frac{1}{M}\sum_i\Theta(|h_i|-\lambda)$,\\
\ \ \ \ \ $x_i^{(n)}=(1-\delta)x_i^{(n-1)}+\delta\frac{1}{\hat{Q}}\mathrm{sign}(h_i)(|h_i|-\lambda)\Theta(|h_i|-\lambda)$\ \ \  for $i=1,\dots,N$.\\
{\it Output:} a compressed vector $\hat{\bm{x}}=\bm{x}^{(\hat{n})}$.\\ \hline
\end{tabular*}
\end{center}
\caption{The procedure of AMP. $\mathrm{sign}(\cdot)$ is the sign function. $\Theta(\cdot)$ is the Heaviside step function.}
\label{AMP}
\end{figure}

\begin{figure}[t]
\begin{tabular}{cc}
\begin{minipage}{0.5\hsize}
\begin{center}
\includegraphics[width=\hsize]{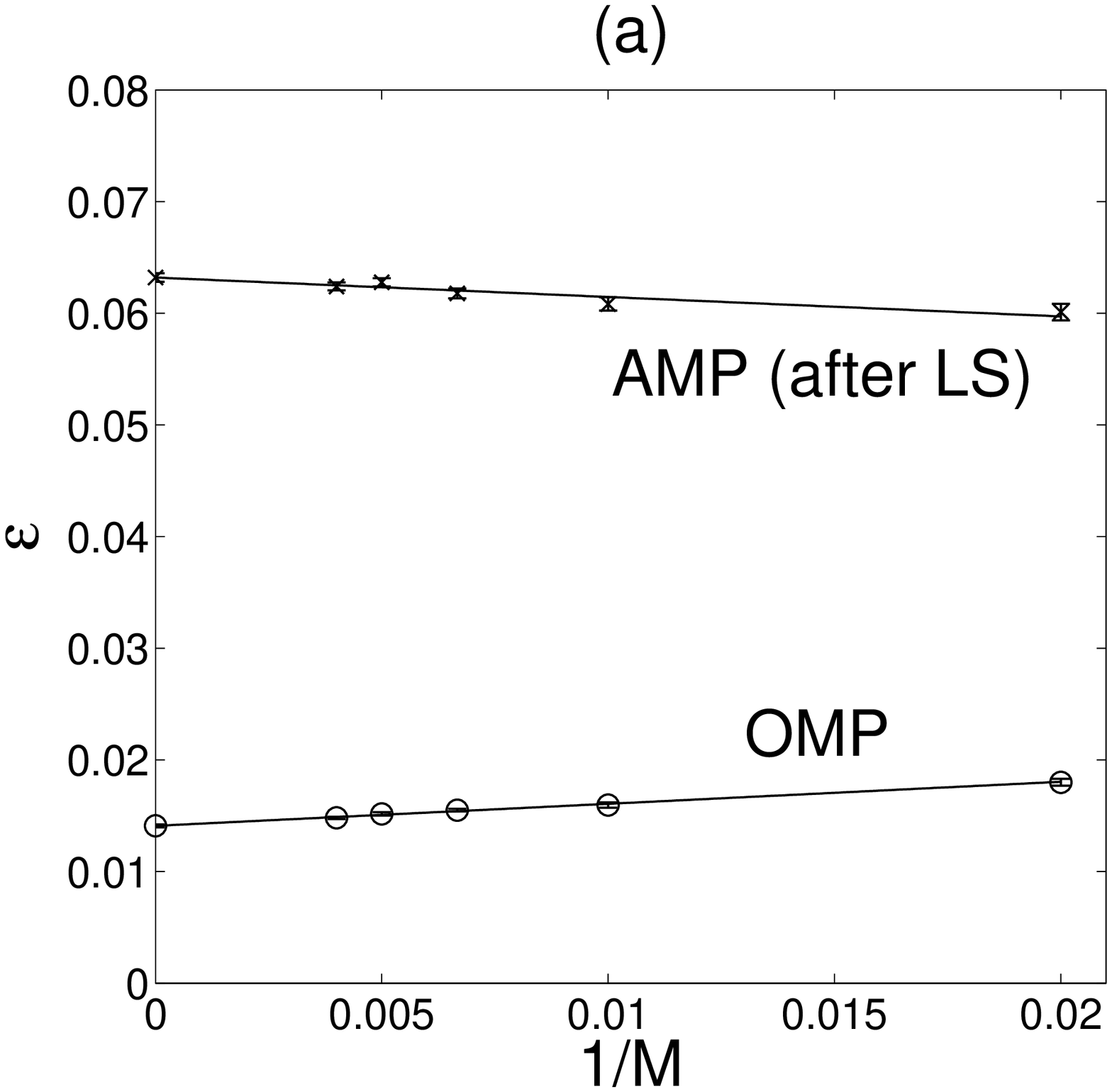}
\end{center}
\end{minipage}
\begin{minipage}{0.5\hsize}
\begin{center}
\includegraphics[width=\hsize]{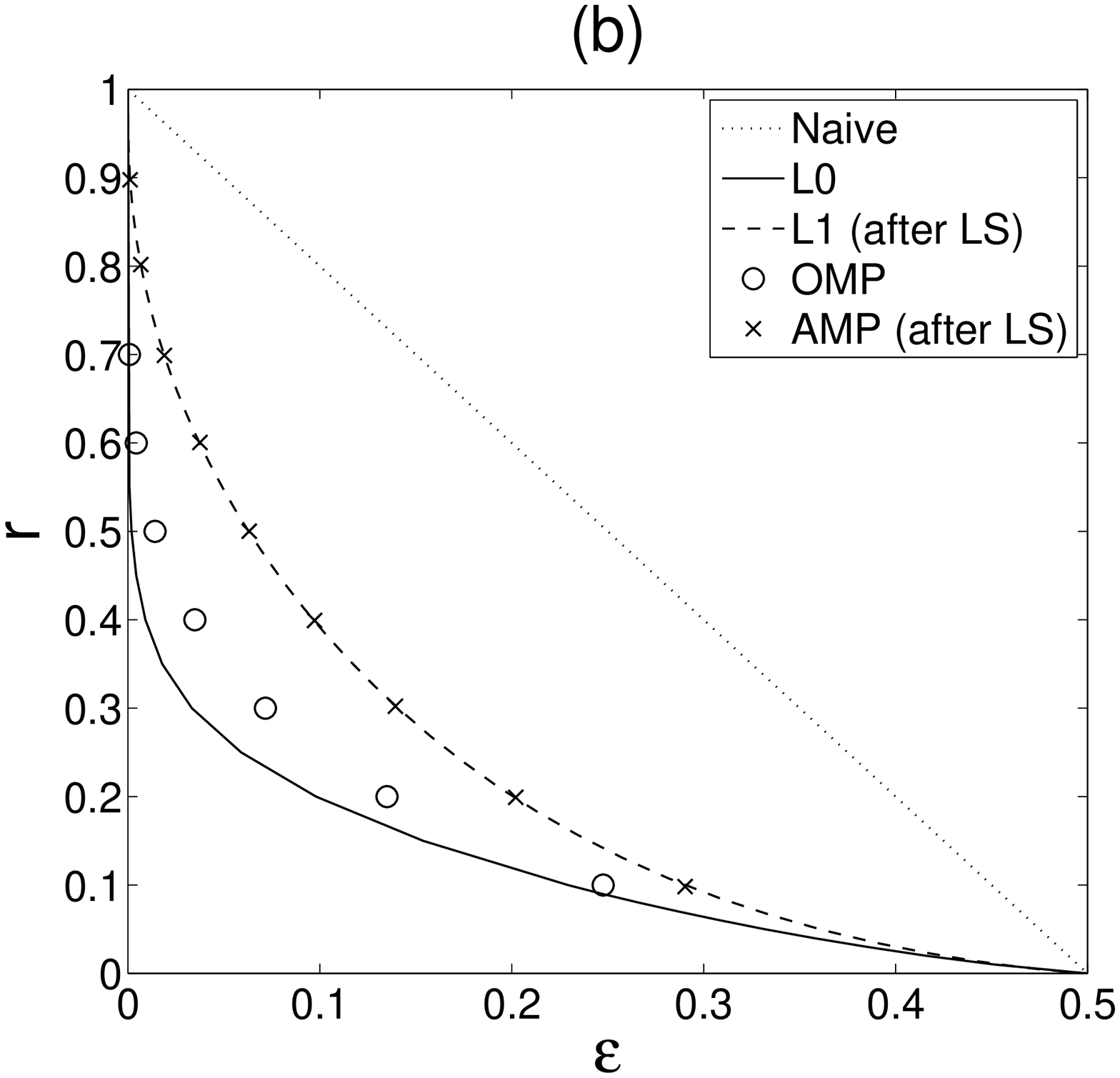}
\end{center}
\end{minipage}
\end{tabular}
\caption{Performances of OMP and AMP in the case with $\sigma_y^2=1$ and $\alpha=0.5$. 
The performance of AMP is evaluated after the method of LS. 
(a) Plots of the numerically evaluated distortions in the case of $r=0.5$.
The extrapolation lines are given by the linear regression.
On the vertical axis, the symbols represent extrapolated values in the $M\to\infty$ limit. 
The lengths of the error bars are comparable to the sizes of symbols. 
In OMP $r$ is set to $r=0.5$, and in AMP $\lambda$ is set to 0.65, so that $r\approx0.5$. 
(b) Trade-off relations in the $M\to\infty$ limit. 
The circles and crosses represent extrapolated values of OMP and AMP, respectively. }
\label{RDalg}
\end{figure}

We evaluate the performances of OMP and AMP when they are employed for sparse approximation with the OCB-based strategy.
We examine the case with $\sigma_y^2=1$ and $\alpha=0.5$.
Figure \ref{RDalg} presents the results of the performance evaluations of OMP and AMP.
Figure \ref{RDalg}(a) shows the results for finite-size systems, namely $M=50,100,\dots,250$, and the extrapolation by the linear regression using 
an asymptotic 
form of $\epsilon\approx a+bM^{-1}$.
The compression rate is set to $r=0.5$ when evaluating OMP, and the regularization coefficient $\lambda$ is set to $0.65$ when evaluating AMP, so that $r\approx0.5$.
We evaluate the performance of AMP based on the distortion after the method of LS.
In figure \ref{RDalg}~(b), we compare the extrapolated performances of OMP and AMP at various rates with the achievable trade-off relation analyzed in section \ref{effect}. 
The AMP result compares well with the ideal performance of the $\ell_1$-based method, while that for OMP does not reach the ideal result of the $\ell_0$-based method.
However, a notable finding is that OMP considerably outperforms the $\ell_1$-based results.
This motivates the exploration of better algorithms for the $\ell_0$-based method, in the context of sparse approximation.
Such exploration is currently under way. 

\subsection{Application to image data}
\begin{figure}[t]
\begin{tabular}{ccc}
\begin{minipage}{0.33\hsize}
\setlength\unitlength{1truecm}
 \begin{picture}(5,5)(0,0)
\put(0,0){\includegraphics[width=\hsize]{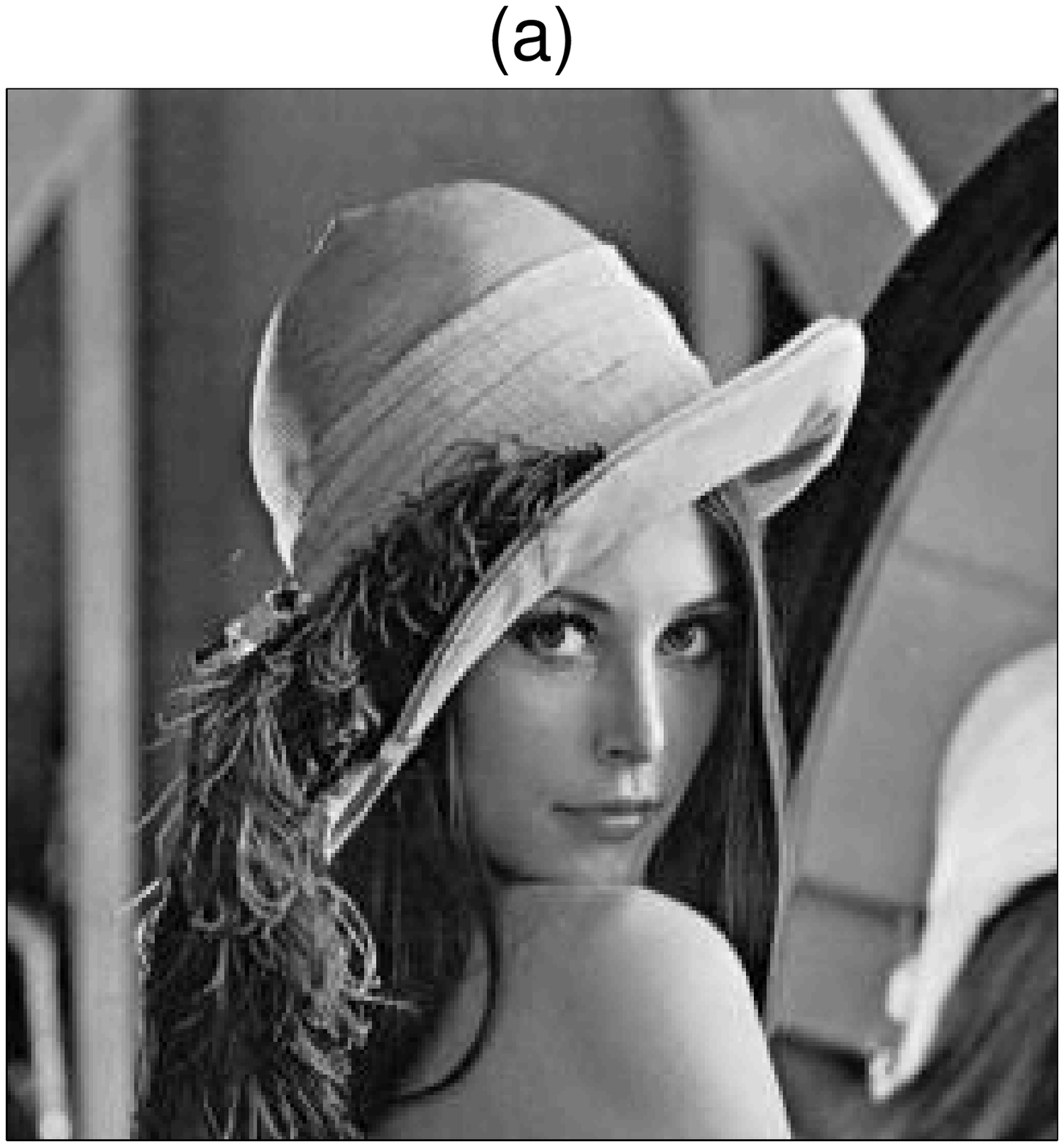}}
\put(.7,0){{\footnotesize \copyright Playboy Enterprises, Inc.}}
\end{picture}
\end{minipage}
\begin{minipage}{0.33\hsize}
\setlength\unitlength{1truecm}
 \begin{picture}(5,5)(0,0)
\put(0,0){\includegraphics[width=\hsize]{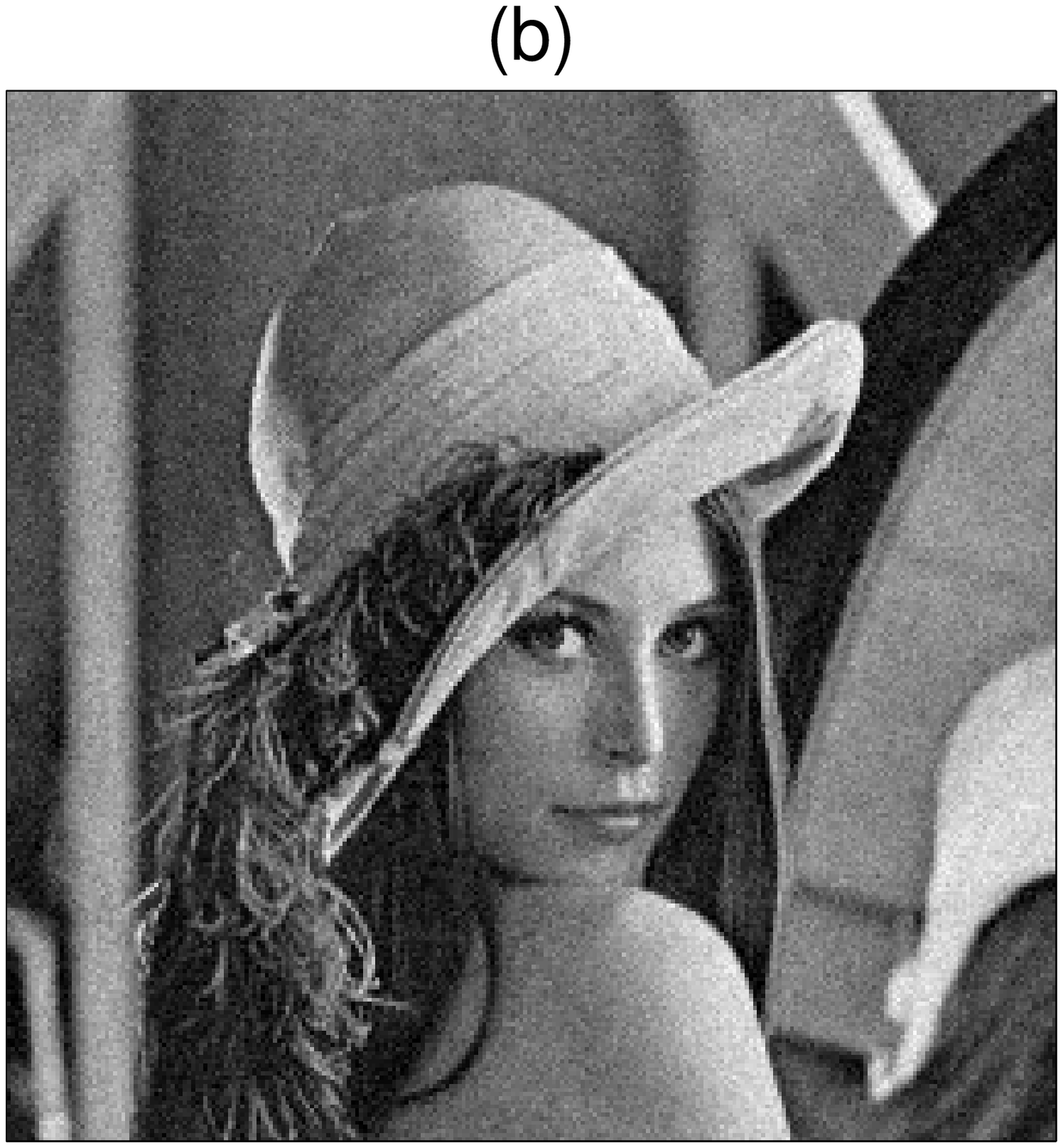}}
\put(.7,0){{\footnotesize \copyright Playboy Enterprises, Inc.}}
\end{picture}
\end{minipage}
\begin{minipage}{0.33\hsize}
\setlength\unitlength{1truecm}
 \begin{picture}(5,5)(0,0)
\put(0,0){\includegraphics[width=\hsize]{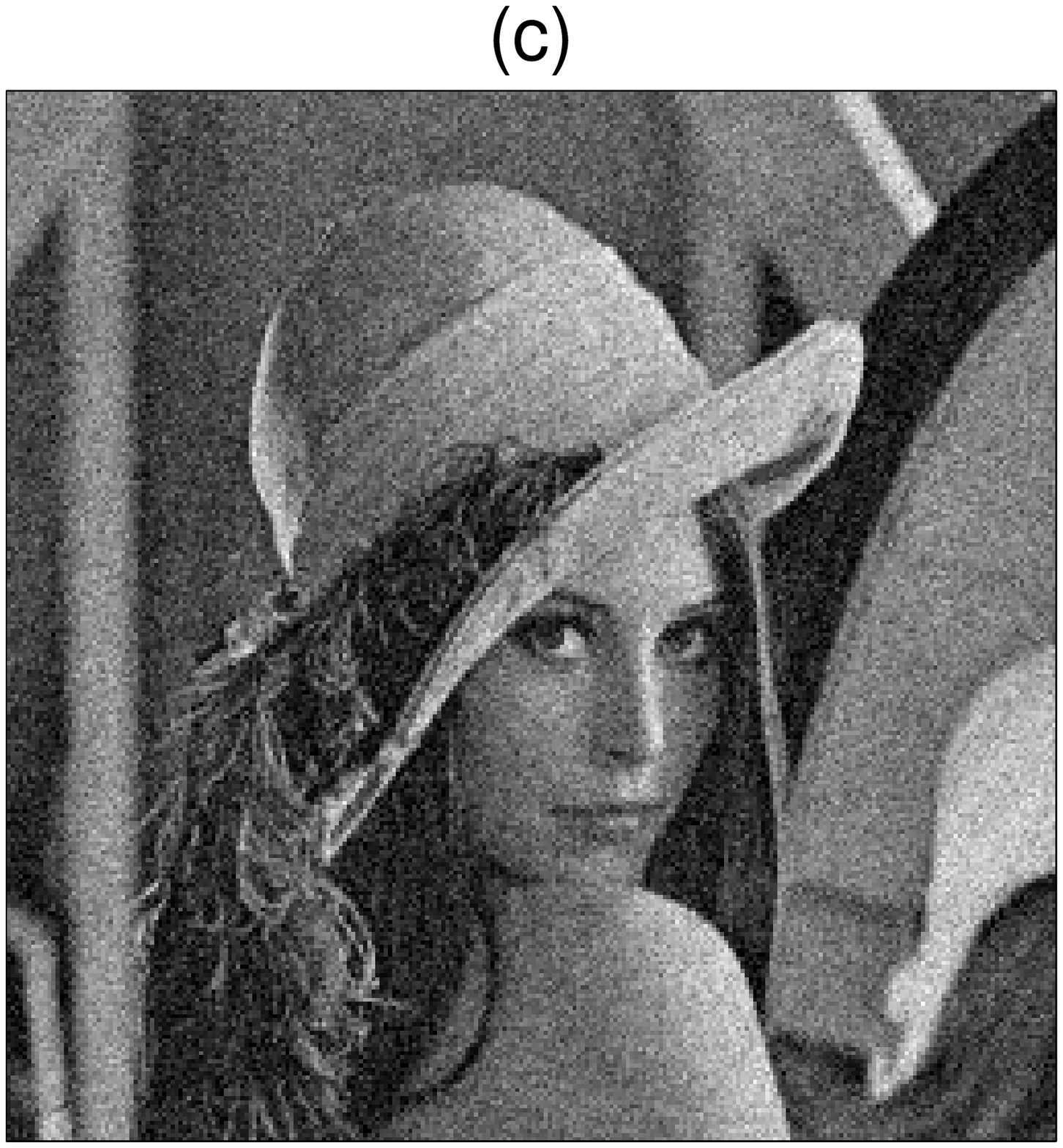}}
\put(.7,0){{\footnotesize \copyright Playboy Enterprises, Inc.}}
\end{picture}
\end{minipage}
\end{tabular}
\caption{Application of sparse approximation with the OCB-based strategy to image data compression. 
The degree of overcompleteness is $\alpha=0.5$. 
(a) Original image data. 
(b) Compressed image data obtained using OMP. 
The compression rate is $r=0.5$. 
PSNR is $28.2$. 
The time required is approximately 55 sec. 
(c) Compressed image data obtained using AMP. 
The regularization coefficient is $\lambda=0.65$, so that $r\approx0.5$.
The AMP-compressed representation is given after the method of LS. 
PSNR is $22.9$. 
The time required is approximately 4.5 sec.
}
\label{Lena}
\end{figure}

We investigate the performance of sparse approximation, when it is applied to a task of image data compression.
We compress image data composed of $256\times256$ pixels.
The experimental procedure of compression is as follows.
First, image data are normalized so as to set the mean and variance to $0$ and $1$, respectively.
Next, $256\times 256$ pixels are randomly permuted, in order to obtain $1024$ column vectors, whose dimension is $64$.
Following these operations, the data can be regarded as random numbers with a mean and variance of 0 and 1, which approximates the properties of the data to the situation which we have already studied theoretically and numerically.
Finally, setting $r=0.5$, we compress each of the column vectors into a compressed vector by using a $64\times 128$ random matrix, namely $\alpha=0.5$.
We examine the performances of OMP and AMP.
When applying AMP, we set the regularization coefficient to $0.65$, so that $r\approx0.5$, and the method of LS is operated after the support estimation by the $\ell_1$-norm regularization.
The results of experiments are presented in figure \ref{Lena}.
Although OMP requires a computational time that is several times larger than that of AMP,
OMP outperforms AMP in terms of appearance and peak signal-to-noise ratio (PSNR), defined as
\begin{eqnarray}
\mathrm{PSNR}=10\log_{10}\frac{255^2}{\frac{1}{N}\sum_{ij}(\hat{I}_{ij}-I_{ij})^2}, 
\end{eqnarray}
where $\bm{I}=\{I_{ij}\}$ and $\hat{\bm{I}}=\{\hat{I}_{ij}\}$ represent an original image and a compressed image, respectively, and $N$ is the number of image pixels.

If the scope of application is limited to image data compression, more convenient bases, such as a discrete wavelet transformation, will achieve much better results in the performance and computational time \cite{Skodras01,Candes08}.
However, in general contexts it is not easy to find a proper basis for sparse approximation in advance. A solution to this problem is to use blind compressed sensing and related techniques such as dictionary learning~\cite{Rubinstein10,Gleichman11,Sakata13}, but the computational costs are rather high. Our OCB-based strategy may overcome this difficulty, because it avoids the learning of the dictionary by preparing many candidates for basis vectors and choosing a suitable combination. Our theoretical analysis and numerical experiments positively support this possibility. 

\section{Conclusion}\label{conclusion}
In the present paper, sparse-data processing has been discussed from the viewpoint of sparse approximation.
We have focused on a strategy of sparse approximation that is based on a random OCB, and have discussed the use of the $\ell_0$- and $\ell_1$-based methods.
We have analyzed the ideal performances of these methods in the large-system limit in a statistical-mechanical manner, which has been validated by numerical simulations on finite-size systems and their extrapolation to the infinite-size limit.
Our results have indicated that the $\ell_0$-based method outperforms the naive and $\ell_1$-based methods in terms of the trade-off relation between the representation distortion and the compression rate. A notable result is that any small distortion is achievable for any finite fixed value of the compression rate, by increasing the degree of overcompleteness, for both the $\ell_0$- and $\ell_1$-based methods. This result allows us to determine both the theoretical limit of the OCB-based strategy and the limit for practical algorithms based on the $\ell_1$ regularization. In addition, it provides a firm basis for the use of the method of LS after the $\ell_1$ regularization, which is frequently applied in related problems such as compressed sensing in practical situations. 

In addition to the ideal performance analyzed in section \ref{effect}, we also investigated the practical performance of our strategy in section \ref{practice}. We evaluated the performances of OMP and AMP as algorithms to approximately perform the $\ell_0$- and $\ell_1$-based methods, respectively.
Our evaluation showed that OMP surpasses both AMP and the exact execution of the $\ell_1$-based method, in terms of the trade-off relation.
This suggests that greedy algorithms are more suitable for sparse approximation using our strategy than convex relaxation algorithms, although there is still room to design more effective greedy algorithms than OMP. We are currently undertaking further research in this direction.

We considered the application of our method to image data compression, as a practical example, and evaluated its performance when OMP and AMP are utilized.
OMP outperforms AMP in appearance and PSNR, although OMP requires a computational time that is several times larger. 
In order to efficiently decrease the computational time of our strategy, it is important to find a proper basis. This suggests the use of some prior knowledge in constructing the overcomplete basis. Some further possibilities, such as combining our methods with dictionary learning, are still open, and would be interesting to address in future work. 


\acknowledgements
This work was supported by JSPS KAKENHI Grant Numbers 13J04920 (YN-O), 26870185 (TO), 25120009 (MO), and 
25120013 (YK).

\appendix
\section{Calculations for the $\ell_0$-based method}
\subsection{Derivation of $\phi_0$}
Based on \BReq{replica_l0}, we define
\begin{eqnarray}
\psi_0(n,\nu,\mu)=
 \frac{1}{M} \ln 
  \lsb
   \lbb
    \Tr{\V{c}} 
      \lb 
        \int \mathrm{d}_{\V{c}}\V{x} \mathrm{e}^{-\frac{1}{2}\frac{\mu}{\nu}||\V{y}-\V{A}(\V{c}\circ \V{x}) ||_2^2   }  
      \rb^{\nu}
   \rbb^n
  \rsb_{\V{y},\V{A}}.
\end{eqnarray}
The cumulant-generating function $\phi_0$ is recovered from $\psi_0$, as $\phi_0(\mu)=\lim_{n,\nu\to0}(1/n)\psi_0(n,\nu,\mu)$. When $(n,\nu)$ are positive integers, we obtain
\begin{eqnarray}
\psi_0(n,\nu,\mu)
=
\frac{1}{M}
\ln
\mathop{\rm Tr}
\limits_{\{\bm{c}^a\}}
\mathop{\rm Tr}\limits_{\{\bm{x}^{a\alpha}\}|\{\bm{c}^{a}\}}
\left[\mathrm{e}^{-\frac{\mu}{2\nu}\sum_{j=1}^{M}\sum_{a=1}^n\sum_{\alpha=1}^{\nu}\left(y_j-\sum_iA_{ji}c_i^ax_i^{a\alpha}\right)^2}
\right]_{\bm{y},\bm{A}}
,
\end{eqnarray}
where $\mathop{\rm Tr}\limits_{\{\bm{c}^a\}}=\prod_{a=1}^n\sum_{\bm{c}^a}\delta(Mr-\sum_ic_i^a)$, and
$\mathop{\rm Tr}\limits_{\{\bm{x}^{a\alpha}\}|\{\bm{c}^a\}}=\prod_{a=1}^n\prod_{\alpha=1}^\nu\int\mathrm{d}_{\bm{c}^a}\bm{x}^{a\alpha}$.
Let us introduce the variables $s_j^{a\alpha}=\sum_iA_{ji}c_i^ax_i^{a\alpha}$ and $Q_{(a\alpha)(b\beta)}=\frac{1}{M}\sum_i(c_i^ax_i^{a\alpha})(c_i^bx_i^{b\beta})$. 
According to the central limit theorem, we regard the variables $\{s_j^{a\alpha}\}$ as random variables that follow a zero-mean multivariate normal distribution, with covariances $[s_j^{a\alpha}s_k^{b\beta}]_{\bm{A}}=\delta_{jk}Q_{(a\alpha)(b\beta)}$. Using these variables, we obtain
\begin{eqnarray}
	\psi_0(n,\nu,\mu)&=&\frac{1}{M}
	\ln
	\mathop{\rm Tr}\limits_{\{\bm{c}^a\}}
	\mathop{\rm Tr}\limits_{ \{\bm{x}^{a\alpha}   \}|\{\bm{c}^a\}}
	\mathop{\rm Tr}\limits_{\bm{Q}}\left(\left[\mathrm{e}^{-\frac{\mu}{2\nu}\sum_a\sum_\alpha(y-s^{a\alpha})^2}\right]_{y,\{s^{a\alpha}\}|\bm{Q}}\right)^M,
\end{eqnarray}
where $\mathop{\rm Tr}\limits_{\bm{Q}}=\prod_{(a\alpha),(b\beta)}\int\mathrm{d}Q_{(a\alpha)(b\beta)}\delta(MQ_{(a\alpha)(b\beta)}-\sum_i(c_i^ax_i^{a\alpha})(c_i^bx_i^{b\beta}))$, and the brackets $\lsb \cdot \rsb_{y,\{s^{a\alpha}\}|\bm{Q}}$ denote the average over $y$ and $s^{a\alpha}$, which is conditioned by the variance $Q_{(a\alpha)(b\beta)}$ as explained above. 

After introducing the Fourier representation of the delta function, $\delta(\cdot)\propto\int\mathrm{d}\tilde{x}\mathrm{e}^{\frac{\tilde{x}}{2}(\cdot)}$, the saddle-point method is employed to obtain
\begin{eqnarray}
&&\psi_0(n,\nu,\mu)
	=\mathop{\rm extr}\limits_{\Theta_0}\Biggl\{\ln\left[\mathrm{e}^{-\frac{\mu}{2\nu}\sum_a\sum_\alpha\left(y-s^{a\alpha}\right)^2}\right]_{y,\{s^{a\alpha}\}|\bm{Q}}\Biggr.
	+\sum_a\frac{\tilde{r}_a}{2}r
	\no \\ &&
	+\sum_{(a\alpha),(b\beta)}\frac{\tilde{Q}_{(a\alpha)(b\beta)}}{2}Q_{(a\alpha)(b\beta)}
	\Biggl.
	+
	\frac{1}{\alpha}\ln\sum_{\{c^a\}}\mathop{\rm Tr}\limits_{\{x^{a\alpha}\}|\{c^a\}}\mathrm{e}^{-\sum_a\frac{\tilde{r}_a}{2}c^a-\sum_{(a\alpha),(b\beta)}\frac{\tilde{Q}_{(a\alpha)(b\beta)}}{2}(c^ax^{a\alpha})(c^bx^{b\beta})}\Biggr\},
\end{eqnarray}
where $\Theta_0=\{\bm{Q},\tilde{\bm{r}},\tilde{\bm{Q}}\}$.
For the extremizer, we search the subspace with $(Q_{(a\alpha)(b\beta)},\tilde{Q}_{(a\alpha)(b\beta)})$ equal to $(Q,\tilde{Q})$ ($a=b,\alpha=\beta$), $(q_1,-\tilde{q}_1)$ ($a=b,\alpha\not=\beta$), or $(q_0,-\tilde{q}_0)$ ($a\not=b$), with $\tilde{r}_a=\tilde{r}$. This is the RS in the present formula of two replica numbers $n$ and $\nu$. Then, we obtain
\begin{eqnarray}
	&&
	\psi_0(n,\nu,\mu)
	=\mathop{\rm extr}\limits_{\tilde{\Theta}_0}\Biggl\{\ln\int\mathrm{D}y\mathrm{D}w\left(\int\mathrm{D}v\left(\int\mathrm{D}u\mathrm{e}^{-\frac{\mu}{2\nu}(\sigma_yy-\sqrt{q_0}w-\sqrt{q_1-q_0}v-\sqrt{Q-q_1}u)^2}\right)^\nu\right)^n\Biggr.
	\nonumber\\ &&
	+\frac{1}{2}n\tilde{r}r+\frac{1}{2}n\nu \tilde{Q}Q-\frac{1}{2}n\nu(\nu-1)\tilde{q}_1q_1-\frac{1}{2}n(n-1)\nu^2\tilde{q}_0q_0
	\nonumber\\ &&
	\Biggl.+\frac{1}{\alpha}\ln\int\mathrm{D}z\left(\int\mathrm{D}t\sum_c\left(\mathop{\rm Tr}\limits_{x|c}\mathrm{e}^{-\frac{\tilde{r}}{2\nu}c-\frac{\tilde{Q}+\tilde{q}_1}{2}cx^2+t\sqrt{\tilde{q}_1-\tilde{q}_0}cx+z\sqrt{\tilde{q}_0}cx}\right)^\nu\right)^n\Biggr\},\label{AppAmain}
\end{eqnarray}
where $\tilde{\Theta}_0=\{Q,q_1,q_0,\tilde{r},\tilde{Q},\tilde{q}_1,\tilde{q}_0\}$ and $\mathop{\rm Tr}\limits_{x|c}=\int\mathrm{d}_cx$. We assume that (\ref{AppAmain}) is true not only for positive integers $(n,\nu)$ but also for real numbers $(n,\nu)$. In taking the limits $(n,\nu)\to(0,0)$, we introduce $\chi=\beta(Q-q_1)$, $q= q_0$, $\hat{r}= \tilde{r}$, $\hat{Q}= \nu(\tilde{Q}+\tilde{q}_1)-\nu^2\tilde{q}_1$, $\hat{\chi}=\nu^2\tilde{q}_1$, and $\hat{q}=\nu^2\tilde{q}_0$, which are assumed to be of the order $O(1)$ in these limits. Following some straightforward calculations, the replica identity is given by
\be
	\phi_0(\mu)=\lim_{n\to0}\lim_{\nu\to0}\frac{1}{n}\psi_0(n,\nu,\mu),
\ee
thus yielding \BReq{phi_0}.

\subsection{The limit $\alpha \to 0$ in the $\ell_0$ case}\Lsec{alphazero_l0}
We examine the behavior of the zero point of entropy, $\epsilon_0$,  in the large-size limit of the basis matrix, $\alpha \to 0$. 
The parameter $\mu$ corresponding to the zero point $\epsilon_0$, $\mu_0$, can be formally written using \BReqs{epsilon(mu)}{s(mu)}, as
\be
&&
\mu_0=-\frac{\hat{\chi}(\mu_0)}{2\phi_0(\mu_0)}
=-\frac{1}{2}\hat{\chi}
\Biggl\{ 
\frac{1}{2}\ln\frac{1+\chi}{1+\chi+\mu_0\Delta}-\frac{1}{2}\frac{\mu_0 (q+\sigma_y^2)}{1+\chi+\mu_0\Delta }
	\Biggr.
	\no\\&& \hspace{-0.5cm}
	+\frac{1}{2}
	\left(
	\hat{r}r+\hat{Q}Q-\frac{\hat{\chi}}{\mu_0}\chi+\hat{q}q
	\right)
	\Biggl.+\frac{1}{\alpha}\int\mathrm{D}z\ln\Biggl(
	1+
	\sqrt{\frac{ \hat{\chi}+\hat{Q} }{ \hat{Q}+\hat{q}}}
	\mathrm{e}^{-\frac{1}{2}\hat{r}+\frac{1}{2}\frac{\hat{q}}{\hat{Q}+\hat{q}}z^2}\Biggr)
\Biggr\}^{-1}.
\Leq{mu_0}
\ee
A numerical calculation indicates the behavior of $\mu_0\to\infty$ as $\alpha\to0$, while $\hat{Q},\hat{q},Q,q,\chi\sim O(1)$ are kept finite.
We will determine the scalings of the relevant variables for $\alpha\to 0$ so as to agree with these observations.
A crucial observation from \BReq{r_l0} is that the factor $Y$ should vanish, in order to cancel the vanishing $\alpha$, yielding
\be
\frac{1}{\alpha }Y \propto \frac{\sqrt{\mu_0}}{\alpha}\mathrm{e}^{-\frac{1}{2}\hat{r} }=O(1) \Rightarrow 
\left\{
\begin{array}{c}
\hat{r}=\tilde{r}-2 \rho \ln \alpha    \\
 \mu_0 \propto \alpha^{2-2\rho }    
\end{array}
\right.,
\Leq{rhat_alphazero}
\ee
where we introduce an exponent $\rho$ controlling the divergence speed of $\hat{r}$ and $\mu_0$. Since we assume the divergence of $\mu_0$, $\rho$ must be larger than unity. The value of $\rho$ is determined by solving \BReq{mu_0} in a self-consistent manner. The scaling of the remaining order parameter $\hat{\chi}$ is determined by
\be
\hat{\chi} \to \frac{\mu_0(\alpha) }{1+\chi}+\frac{q+\sigma_y^2}{\Delta^2} \to \infty. 
\Leq{chihat_reduced}
\ee
Now, we know all of the scalings of the order parameters, and can reduce \BReq{mu_0} to the dominant part, as 
\be
\mu_0\approx \frac{\frac{\mu_0}{1+\chi} +\frac{q+\sigma_y^2}{\Delta^2}}{2\rho r\ln \alpha+\ln (1+\chi+\mu_0 \Delta)}.
\Leq{mu_0-reduced}
\ee
By solving this in the leading scaling, we obtain
\be
\rho=\frac{1}{1-r},~
\mu_{0} \approx \frac{\mathrm{e}^{(1+\chi)^{-1} } }{\Delta} \alpha^{-2\frac{r}{1-r}}+O(1).
\Leq{mu_0-sol}
\ee
By inserting \BReqs{chihat_reduced}{mu_0-sol} into \BReq{epsilon(mu)}, we get \BReq{epsilon_0-alphazero}

\section{Some calculations for the $\ell_1$-based methods}

\subsection{Derivations of $f_1$ and $\epsilon^{\rm {LS}}_1$}
Based on \BReq{replica_l1+PI}, we introduce
\begin{eqnarray}
&&
\psi_1(n,\nu,\beta,\mu,\kappa)
=
\frac{1}{M}
\ln
\left[
Z^{n-1}_1(\mu,\kappa|\bm{y},\bm{A})
\int\mathrm{d}\bm{\xi} 
 \mathrm{e}^{-\mu(\mathcal{H}_1(\bm{\xi}|\bm{y},\bm{A})+\kappa||\bm{\xi}||_0)}
\lb 
\int\mathrm{d}_{|\V{\xi}|_0}\bm{x}
\mathrm{e}^{-\frac{\beta}{2}||\bm{y}-\bm{A}(|\bm{\xi}|_0\circ\bm{x})||_2^2}
\rb^\nu
\right]_{\bm{y},\bm{A}}.
\end{eqnarray}
By calculating this in the case of positive integers $(n,\nu)$, we obtain
\begin{eqnarray}
&&
\psi_1(n,\nu,\beta,\mu,\kappa)
=\frac{1}{M}\ln
\mathop{\rm Tr}\limits_{\{\bm{\xi}^a\}}
\mathop{\rm Tr}\limits_{\{\bm{x}^\alpha\}}
\left(\left[\mathrm{e}^{-\frac{\mu}{2}\sum_a(y_j-\sum_i A_{ji}\xi_i^a)^2
-\frac{\beta}{2}\sum_\alpha(y_j-\sum_i A_{ji} |\xi_i^1|_0x_i^\alpha)^2}\right]_{\bm{y},\bm{A}}\right)^M,
\end{eqnarray}
where $\mathop{\rm Tr}\limits_{\{\bm{\xi}^a\}}=\prod_{a=1}^n\int\mathrm{d}\bm{\xi}^a\mathrm{e}^{-\mu(\lambda\sum_i|\xi_i^a|+\kappa\sum_i|\xi_i^a|_0)}$, and $\mathop{\rm Tr}\limits_{\{\bm{x}^\alpha\}  }=\prod_{\alpha=1}^\nu\int\mathrm{d}_{|\bm{\xi}^1|_0}\bm{x}^{\alpha}$.
Let us introduce the variables
$s_j'^a=\sum_i A_{ji}\xi_i^a$, $s_j^\alpha=\sum_i A_{ji}|\xi_i^1|x_i^\alpha$, 
$P_{ab}=\frac{1}{M}\sum_i\xi_i^a\xi_i^b$, 
$C_{\alpha a}=\frac{1}{M}\sum_i(|\xi_i^1|_0x_i^\alpha)\xi_i^a$, 
and $Q_{\alpha\beta}=\frac{1}{M}\sum_i(|\xi_i^1|_0x_i^\alpha)(|\xi_i^1|_0x_i^\beta)$.
As in the $\ell_0$ case, we can rewrite the variables $\{s_j'^a,s_j^\alpha\}$ as random variables from a zero-mean multivariate normal distribution, with the covariances $[s_j'^as_k'^b]_{\bm{A}}=\delta_{jk}P_{ab}$, 
$[s'^a_j s^\alpha_k]_{\bm{A}}=\delta_{jk}C_{a\alpha}$, 
and $[s_j^\alpha s_k^\beta]_{\bm{A}}=\delta_{jk}Q_{\alpha\beta}$.
The application of the central limit theorem here is justified by the nonzeroness of compression rate $r$ shown in figure \ref{RD}~(b) derived from (\ref{eq:r_l1}).
Using these variables, we obtain
\begin{eqnarray}
&&
\psi_1(n,\nu,\beta,\mu,\kappa)
\no \\ &&
=\frac{1}{M}\ln
\mathop{\rm Tr}\limits_{\{\bm{\xi}^a\}}
\mathop{\rm Tr}\limits_{\{\bm{x}^\alpha\}}
	\mathop{\rm Tr}\limits_{\bm{P}}\mathop{\rm Tr}\limits_{\bm{C}}
\mathop{\rm Tr}\limits_{\bm{Q}}\left(\left[\mathrm{e}^{-\frac{\mu}{2}\sum_a(y_j-s_j'^a)^2-\frac{\beta}{2}\sum_\alpha(y_j-s_j^\alpha)^2}\right]_{y_j,\{s_j'^a,s_j^\alpha\}|\bm{P},\bm{C},
\bm{Q}}\right)^M,
\end{eqnarray}
where $\mathop{\rm Tr}\limits_{\bm{P}}=\prod_{a,b}\int\mathrm{d}P_{ab}\delta(MP_{ab}-\sum_i\xi_i^a\xi_i^b)$, $\mathop{\rm Tr}\limits_{\bm{C}}=\prod_{a,\alpha}\int\mathrm{d}C_{\alpha a}\delta(MC_{\alpha a}-\sum_i(|\xi_i^1|_0x_i^\alpha)\xi_i^a)$, 
and $\mathop{\rm Tr}\limits_{\bm{Q}}=\prod_{\alpha,\beta}\int\mathrm{d}Q_{\alpha\beta}\delta(MQ_{\alpha\beta}-\sum_i(|\xi_i^1|_0x_i^\alpha)(|\xi_i^1|_0x_i^\beta))$.
After introducing the Fourier representation of the delta function, the saddle-point method is employed, to obtain
\begin{eqnarray}
&&
\psi_1(n,\nu,\beta,\mu,\kappa)
=\mathop{\rm extr}\limits_{\Theta_1}\Biggl\{\ln\left[\mathrm{e}^{-\frac{\mu}{2}\sum_a(y-s'^a)^2-\frac{\beta}{2}\sum_\alpha(y-s^\alpha)^2}\right]_{y,\{s'^a,s^\alpha\}|\bm{P},\bm{C},\bm{Q}}\Biggr.\nonumber\\
	&&+\sum_{a,b}\frac{\tilde{P}_{ab}}{2}P_{ab}+\sum_{a,\alpha}\tilde{C}_{a\alpha}C_{a\alpha}
+\sum_{\alpha,\beta}\frac{\tilde{Q}_{\alpha\beta}}{2}Q_{\alpha\beta}\nonumber\\
	&&\Biggl.+\frac{1}{\alpha}\ln\mathop{\rm Tr}\limits_{\{\xi^a\}}\mathop{\rm Tr}\limits_{\{x^\alpha\}}\mathrm{e}^{-\sum_{a,b}\frac{\tilde{P}_{ab}}{2}\xi^a\xi^b-\sum_{a,\alpha}\tilde{C}_{a\alpha}\xi^a(|\xi^1|_0x^\alpha)-\sum_{\alpha,\beta}\frac{\tilde{Q}_{\alpha\beta}}{2}(|\xi^1|_0x^\alpha)(|\xi^1|_0x^\beta)}\Biggr\},
\end{eqnarray}
where $\Theta_1=\{\bm{P},\bm{C},\bm{Q},\tilde{\bm{P}},\tilde{\bm{C}},\tilde{\bm{Q}}\}$.
For the extremizer, we search the subspace with $(P_{ab},\tilde{P}_{ab})$ equal to $(P,\tilde{P})$ ($a=b$) or $(p,-\tilde{p})$ ($a\not=b$); $(C_{a\alpha},\tilde{C}_{a\alpha})$ equal to $(C,-\tilde{C})$ ($a=1$) or $(c,-\tilde{c})$ ($a\not=1$); and $(Q_{\alpha\beta},\tilde{Q}_{\alpha\beta})$ equal to $(Q,\tilde{Q})$ ($\alpha=\beta$) or $(q,-\tilde{q})$ ($\alpha\not=\beta$). This is the RS assumption for the present case. Thus, we obtain
\begin{eqnarray}
&&
\psi_1(n,\nu,\beta,\mu,\kappa)
=\mathop{\rm extr}\limits_{\tilde{\Theta}^{\rm LS}_1,\tilde{\Theta}_1}\Biggl\{\ln\int\mathrm{D}y\mathrm{D}z\mathrm{D}w\left(\int\mathrm{D}v\mathrm{e}^{-\frac{\mu}{2}(\sigma_yy-\sqrt{p}w-\sqrt{P-p}v)^2}\right)^{n-1}\nonumber\\
	&&\times\int\mathrm{D}v\mathrm{e}^{-\frac{\mu}{2}(\sigma_yy-\sqrt{p}w-\sqrt{P-p}v)^2}
	\left(\int\mathrm{D}u\mathrm{e}^{-\frac{\beta}{2}(\sigma_yy-\frac{c}{\sqrt{p}}w-\frac{C-c}{\sqrt{P-p}}v-\sqrt{q-\frac{(C-c)^2}{P-p}-\frac{c^2}{p}}z-\sqrt{Q-q}u)^2}\right)^\nu\Biggr.\nonumber\\
	&&+\frac{1}{2}n\tilde{P}P-\frac{1}{2}n(n-1)\tilde{p}p-\nu\tilde{C}C-(n-1)\nu\tilde{c}c+\frac{1}{2}\nu\tilde{Q}Q-\frac{1}{2}\nu(\nu-1)\tilde{q}q\nonumber\\
	&&\Biggl.+\frac{1}{\alpha}\ln\int\mathrm{D}z\mathrm{D}w\mathrm{D}v\mathrm{D}u\left(\mathop{\rm Tr}\limits_{\xi}\mathrm{e}^{-\frac{\tilde{P}+\tilde{p}}{2}\xi^2+(u\sqrt{\tilde{p}-\tilde{c}}+v\sqrt{\tilde{c}})\xi}\right)^{n-1}\nonumber\\
	&&\times\mathop{\rm Tr}\limits_\xi\mathrm{e}^{-\frac{\tilde{P}+\tilde{p}+\tilde{C}-\tilde{c}}{2}\xi^2+(u\sqrt{\tilde{p}-\tilde{c}}+w\sqrt{\tilde{C}-\tilde{c}}+v\sqrt{\tilde{c}})\xi}\nonumber\\
	&&\Biggl.\times\left(\mathop{\rm Tr}\limits_{x}\mathrm{e}^{-\frac{\tilde{Q}+\tilde{q}+\tilde{C}-\tilde{c}}{2}(|\xi|_0x)^2+(z\sqrt{\tilde{q}-\tilde{c}}+w\sqrt{\tilde{C}-\tilde{c}}+v\sqrt{\tilde{c}})|\xi|_0x}\right)^\nu\Biggr\},\label{AppCmain}
\end{eqnarray}
where $\tilde{\Theta}^{\rm LS}_1=\{C,c,Q,q,\tilde{C},\tilde{c},\tilde{Q},\tilde{q}\}$ and $\tilde{\Theta}_1=\{P,p,\tilde{P},\tilde{p}\}$. 

The free-energy density $f_1$ is now derived as 
\be
&&
f_1(\mu,\kappa)=-\lim_{n\to 0}\lim_{\nu\to 0}\frac{1}{\mu n}\psi_1(n,\nu,\beta,\mu,\kappa)\\
&&=\mathop{\rm extr}\limits_{\tilde{\Theta}_1}\left\{\frac{1}{2\mu}\ln(1+\mu(P-p))+\frac{1}{2}\frac{P+\sigma_y^2}{1+\mu(P-p)}-\frac{1}{2\mu}(\tilde{P}P+\tilde{p}p)\right.\nonumber\\
&&\left.-\frac{1}{\mu\alpha}\int\mathrm{D}v\ln\mathop{\rm Tr}\limits_{\xi}\mathrm{e}^{-\frac{\tilde{P}+\tilde{p}}{2}\xi^2+v\sqrt{\tilde{p}}\xi}\right\}.
\label{AppBresult}
\ee
In the limit $\mu \to \infty$, we introduce $\chi_p=\mu(P-p)$, $\hat{P}=\mu^{-1}(\tilde{P}+\tilde{p})$, and $\hat{\chi}_p=\mu^{-2}\tilde{p}$, which are assumed to be of the order $O(1)$. Taking the $\mu \to \infty$ limit in (\ref{AppBresult}) leads to (\ref{l1preresult}). 

On the other hand, in order to evaluate $\epsilon_1^{\rm LS}$, in addition to $\hat{\Theta}_1={P,\chi_p,\hat{P},\hat{\chi_p}}$, in taking the limit $\mu\to \infty$ we define the parameters $\chi_c=\beta(C-c)$, $\chi_q=\beta(Q-q)$, $\hat{C}=\beta^{-1}(\tilde{C}+\tilde{c})$, $\hat{\chi}_c=\beta^{-2}\tilde{c}$, $\hat{Q}=\beta^{-1}(\tilde{Q}+\tilde{q})$, and $\hat{\chi}_q=\beta^{-2}\tilde{q}$, which are assumed to be of the order $O(1)$. Then, through the formula
\be
\epsilon_{1}^{\rm LS}= \lim_{\beta\to\infty}\lim_{\mu \to \infty} \lim_{n\to 0}\lim_{\nu\to 0}-\frac{1}{\beta\nu}\psi_1(n,\nu,\beta,\mu,0),
\ee
we obtain \BReq{epsilon_1^PI}.

\subsection{The limit $\alpha \to 0$ in the $\ell_1$ case}\Lsec{alphazero_l1}
The EOSs \NReq{EOSs_l1} show that in the limit $\alpha \to 0$ we have
\be
\chi_p,\hat{P},\hat{\chi}_p=O(1).
\ee
From \BReq{r_l1} and the asymptotic formula of the complementary error function $\rm{erfc}(\cdot)$, we see in the limit $\alpha \to 0$ we have
\be
\frac{\mathrm{e}^{-\frac{1}{2}\theta^2  } }{ \alpha \sqrt{2\pi} \theta }=O(1),
\Rightarrow \theta=O(\sqrt{ | \ln \alpha | } )\to \infty ,
\Leq{theta_limit-L1}
\ee
which is realized by controlling $\lambda$ as $O(\sqrt{|\ln\alpha|})$.
Using these scalings, and the asymptotic expansion of the complementary error function for large $\theta$ in \BReq{P_l1}, we obtain
\be
P=O(\theta^{-2})=O(|\ln \alpha |^{-1}) \to 0. 
\ee 
By inserting these scalings into \BReq{epsilon_l1}, we obtain \BReq{epsilon_1-alphazero}. 

The asymptotic form of $\epsilon_1^{\rm LS}$ can be similarly obtained. Following some lengthy but straightforward calculations, we obtain
\subbe
\be
&&
\hat{\chi}_q=O(|\ln \alpha|^{-1} ) \to 0, 
\\ &&
\hat{Q}=O(1),
\\ &&
\hat{\chi}_c=O( |\ln \alpha|^{-1} ) \to 0, 
\\ &&
\hat{C}=O(1), 
\\ &&
\chi_q=O(1),
\\ &&
Q=O(|\ln \alpha|^{-1} ) \to 0, 
\\ &&
\chi_c=O(1), 
\\ &&
C=O(|\ln \alpha|^{-1} ) \to 0.
\ee
\subee
By substituting these scalings into \BReq{epsilon_1^PI-simples}, we obtain \BReq{epsilon_1^PI-alphazero}.


\end{document}